\documentclass[aps,prx,floatfix,twocolumn,superscriptaddress,10pt, nofootinbib]{revtex4-2}

\usepackage{amsmath,amssymb,amsfonts}
\usepackage{bm}
\usepackage{subcaption}
\usepackage{graphicx}
\usepackage{hyperref}

\usepackage[dvipsnames]{xcolor}
\usepackage[most]{tcolorbox}

\definecolor{contbg}{RGB}{255,240,210} 
\definecolor{discbg}{RGB}{215,230,255} 
\definecolor{conttitle}{RGB}{210,130,30} 
\definecolor{disctitle}{RGB}{40,85,175}  

\begin{document}

\title{Discontinuous BBP transitions}

\author{Dario Bocchi}
\affiliation{Physics Department, University of Rome “La Sapienza”, Piazzale Aldo Moro 5, 00185 Rome, Italy}
\affiliation{Institute of Nanotechnology, CNR-NANOTEC, Piazzale Aldo Moro 5, 00185 Rome, Italy}

\author{Giulio Biroli}
\affiliation{Laboratoire de Physique de l’École normale supérieure, ENS, Université PSL,
CNRS, Sorbonne Université, Université de Paris, F-75005 Paris, France}

\author{Chiara Cammarota}
\affiliation{Physics Department, University of Rome “La Sapienza”, Piazzale Aldo Moro 5, 00185 Rome, Italy}

\author{Federico Ricci-Tersenghi}
\affiliation{Physics Department, University of Rome “La Sapienza”, Piazzale Aldo Moro 5, 00185 Rome, Italy}
\affiliation{Institute of Nanotechnology, CNR-NANOTEC, Piazzale Aldo Moro 5, 00185 Rome, Italy}
\affiliation{INFN -- Sezione di Roma 1, Piazzale Aldo Moro 5, 00185 Rome, Italy}

\begin{abstract}
The Baik–Ben Arous–Péché (BBP) transition sets fundamental limits for detecting low-rank structure in noisy high-dimensional data and underlies a wide range of spectral methods in many fields from physics to statistics and data sciences. In standard settings, this transition is continuous, implying that signal recovery emerges gradually above a sharp threshold.
We show that BBP transitions can instead be discontinuous in very general settings and provide a full theory of this phenomenon. When the eigenvalue density vanishes faster than linearly at the spectral edge, the overlap between the leading eigenvector and the signal jumps discontinuously at the critical point.
We study this mechanism in deformed Gaussian and reweighted Wishart ensembles.
We analyze in detail the finite-size effects, which play a central and qualitatively new role in the discontinuous BBP transition.
Unlike the continuous BBP transition, we establish the existence of an extended pre-critical region where informative eigenvectors emerge well before the asymptotic threshold. 
The main consequence---and difference from the continuous BBP transition---is that signal recovery can occur at significantly lower signal-to-noise ratio and it is accompanied by strong sample-to-sample variability.
Our results show the relevance and the novelty of the discontinuous BBP transition, and highlight the practical implications for signal detection.

\end{abstract}

\maketitle

\section{Introduction} 
Random matrix theory provides a powerful framework to describe the interplay between structure and noise in complex systems with many degrees of freedom. A central paradigm in this context is the  BBP transition, which was first found for disordered magnetic systems by Edwards and Jones \cite{edwards1976eigenvalue}, then in the case of unsupervised learning by Watkin and Nadal \cite{watkin1994optimal}, and more recently in statistics and probability theory by Baik, Ben Arous, P\'ech\'e \cite{baik2005phase}.\\
This phenomenon consists in the emergence of an outlier eigenvalue in a large random matrix subject to a finite-rank perturbation. At this transition, an eigenvalue separates from the spectral bulk and the associated eigenvector becomes correlated with the perturbation, marking the onset of signal detectability.
It is a mechanism that plays a crucial role across a broad range of fields. In high-dimensional statistics, it governs the behavior of principal component analysis and spiked covariance models \cite{paul2007asymptotics}. In signal processing \cite{perry2018optimality} and machine learning \cite{mergny2024spectral}, it sets fundamental limits for detecting low-rank structure in noisy data. BBP transitions arise in many other fields: statistical inference \cite{montanari2015limitation}, network science \cite{lee2022phase}, biology \cite{fraboul2021artificial}, disordered physical systems \cite{krajenbrink2021tilted} and finance \cite{potters2020first}. In all these cases, they delimit phases of detectability and encode sharp thresholds between noise-dominated and signal-dominated regimes. The ubiquity of this phenomenon has established the BBP transition as a cornerstone in data sciences. 

A key feature common to all previously studied instances is that the transition is continuous: the overlap between the leading eigenvector and the signal vanishes at criticality and grows smoothly above the threshold. This behavior is closely tied to the universal square-root decay of the eigenvalue density at spectral edges in standard random matrix ensembles.

The aim of this work is to show that BBP transitions can be -- and are in many cases -- {\it discontinuous}, in the sense that the eigenvector overlap exhibits a finite jump at the critical point. In this regime, signal detectability emerges abruptly rather than continuously, defining a distinct universality class of spectral transitions.

As we shall discuss, the main ingredient to have a discontinuous BBP is that the perturbed random matrix has a density of eigenvalues that vanishes as $(\lambda_+-\lambda)^{a-1}$ with $a>2$ at the right edge $\lambda_+$ (without loss of generality we consider the case in which the rank one perturbation is positive and hence the BBP eigenvalue pops out at the right edge). Although this situation is likely to arise in a broad range of contexts, it has only recently come into focus in random matrix theory \cite{lee2016extremal, potters2020first, bouchbinder2021low, franz2022delocalization}, and its complete analysis and general implications have remained largely unexplored.

In this work, we provide a full theory of discontinuous BBP transitions by focusing on two general classes of random matrices to which a rank-one perturbation is added to induce the BBP transition. The first has been recently studied in random matrix theory \cite{lee2013local, lee2016extremal} and in physics \cite{krajenbrink2021tilted, bouchbinder2021low, franz2022delocalization}. It corresponds to a deformed GOE matrix, i.e., a GOE matrix to which is added a diagonal matrix with elements distributed following a certain continuous non-singular law. The second one corresponds to weighted Wishart random matrices in which the sum of the rank-one terms is weighted by random coefficients. Such matrices arise in statistics and machine learning \cite{mondelli2018fundamental}.

In both cases, we first establish the condition for the existence of a discontinuous BBP transition and then analyze in detail the finite-size effects at large but finite problem size $N$. In the discontinuous regime, these effects lead to an extended pre-critical region in which an outlier and a finite overlap persist well below the asymptotic threshold. As a consequence, signal recovery can occur at significantly lower signal-to-noise ratios in finite systems than predicted by the asymptotic theory.

Beyond this shift in thresholds, discontinuous BBP transitions exhibit a qualitatively different phenomenology. The abrupt onset of the overlap is accompanied by pronounced sample-to-sample fluctuations, leading to strong variability in detection performance across realizations. These features fundamentally modify the behavior of spectral methods in realistic settings, where system sizes are finite and fluctuations cannot be neglected.

Taken together, our results highlight the relevance, the novelty and the practical implications of the discontinuous BBP transition. They highlight a new mechanism for abrupt signal emergence and detectability thresholds with applications ranging from statistical inference to machine learning and complex physical models.

\section{Spectral Theory of BBP Transitions}
\label{sec:spectral}

\begin{figure*}[t]
\centering

\begin{tcolorbox}[
    colback=contbg,
    colframe=contbg,
    width=\textwidth,
    arc=2mm,
    boxrule=0pt,
    left=2mm,right=2mm,top=3mm,bottom=2mm
]
\centering
{\color{conttitle}\textbf{Continuous BBP transition}}\\[3mm]

\begin{subfigure}{0.32\textwidth}
    \centering
    \includegraphics[width=\linewidth]{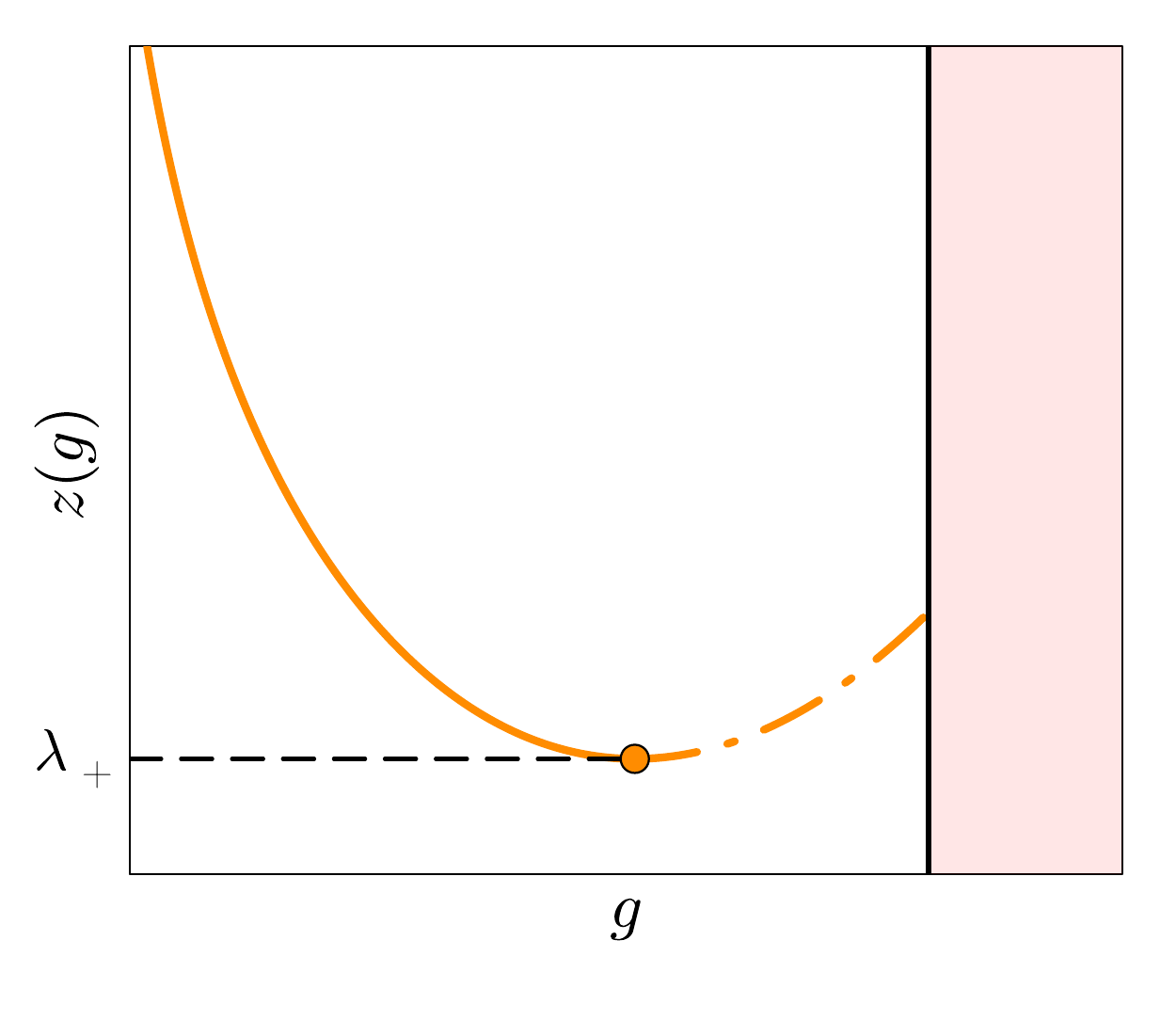}
    \caption*{\small Inverse resolvent $z(g)$}
\end{subfigure}
\hfill
\begin{subfigure}{0.32\textwidth}
    \centering
    \includegraphics[width=\linewidth]{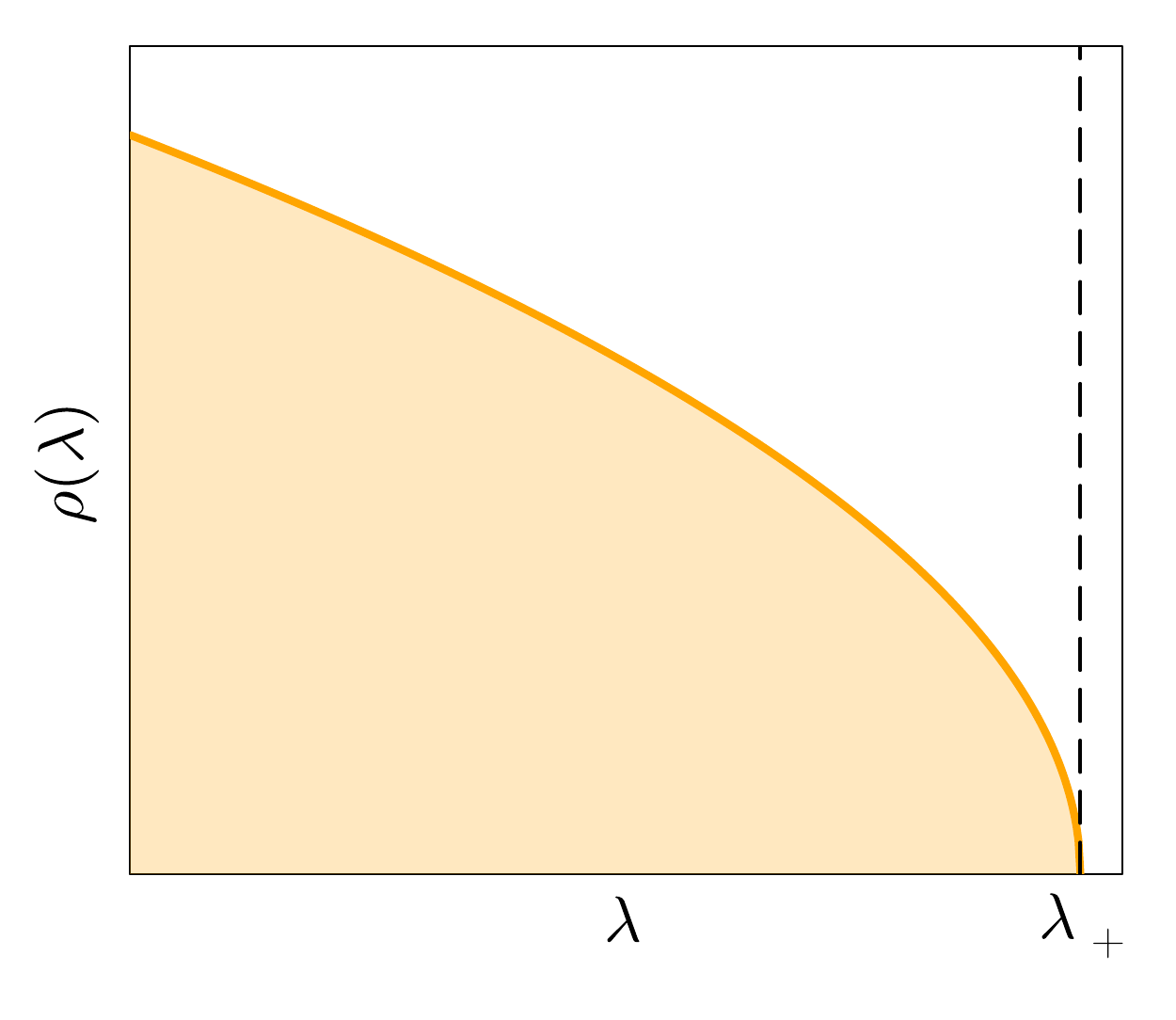}
    \caption*{\small Density $\rho(\lambda)$}
\end{subfigure}
\hfill
\begin{subfigure}{0.32\textwidth}
    \centering
    \includegraphics[width=\linewidth]{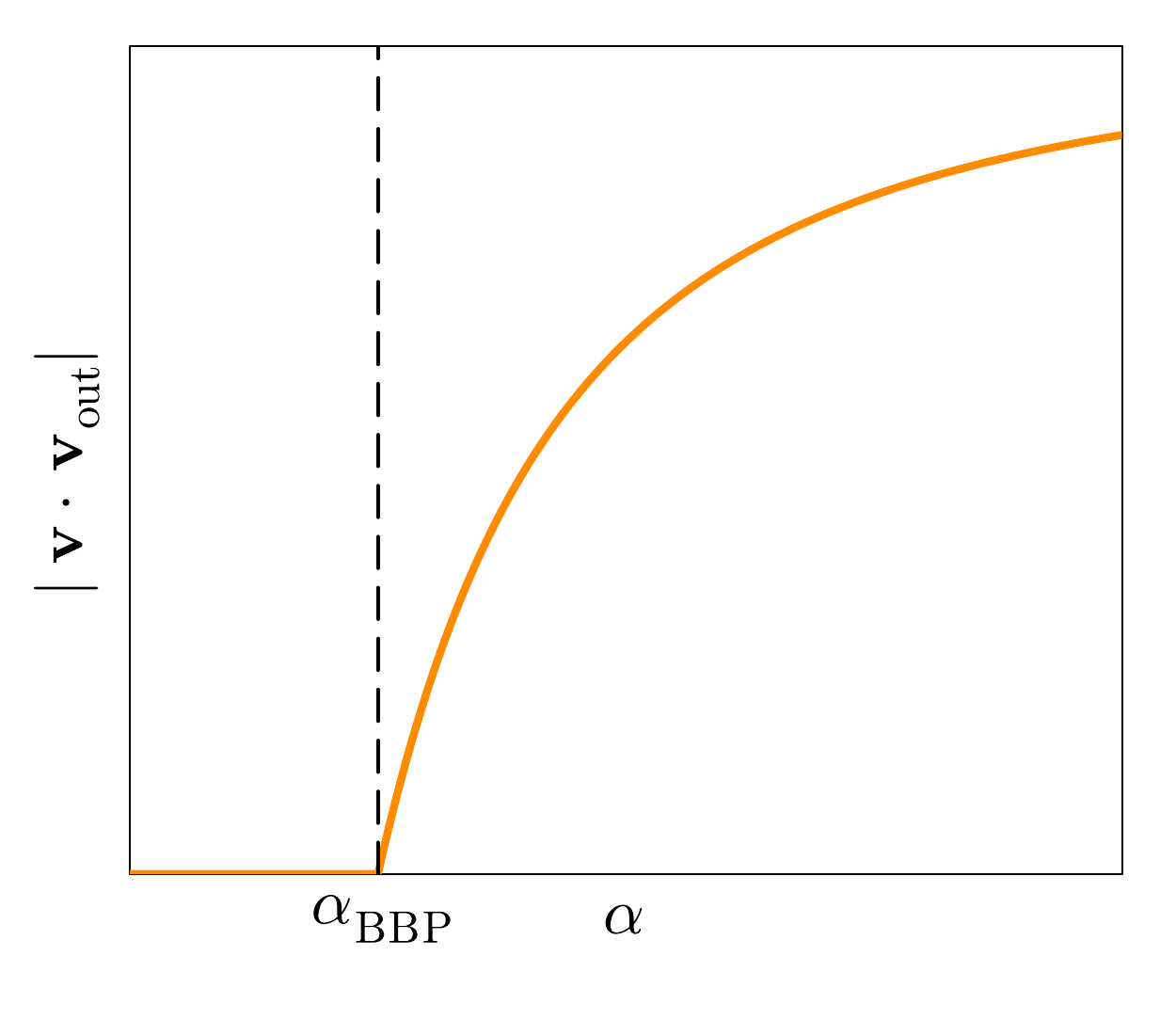}
    \caption*{\small Overlap $|\mathbf{v}\!\cdot\!\mathbf{v}_{\mathrm{out}}|$}
\end{subfigure}
\end{tcolorbox}

\vspace{4mm}

\begin{tcolorbox}[
    colback=discbg,
    colframe=discbg,
    width=\textwidth,
    arc=2mm,
    boxrule=0pt,
    left=2mm,right=2mm,top=3mm,bottom=2mm
]
\centering
{\color{disctitle}\textbf{Discontinuous BBP transition}}\\[3mm]

\begin{subfigure}{0.32\textwidth}
    \centering
    \includegraphics[width=\linewidth]{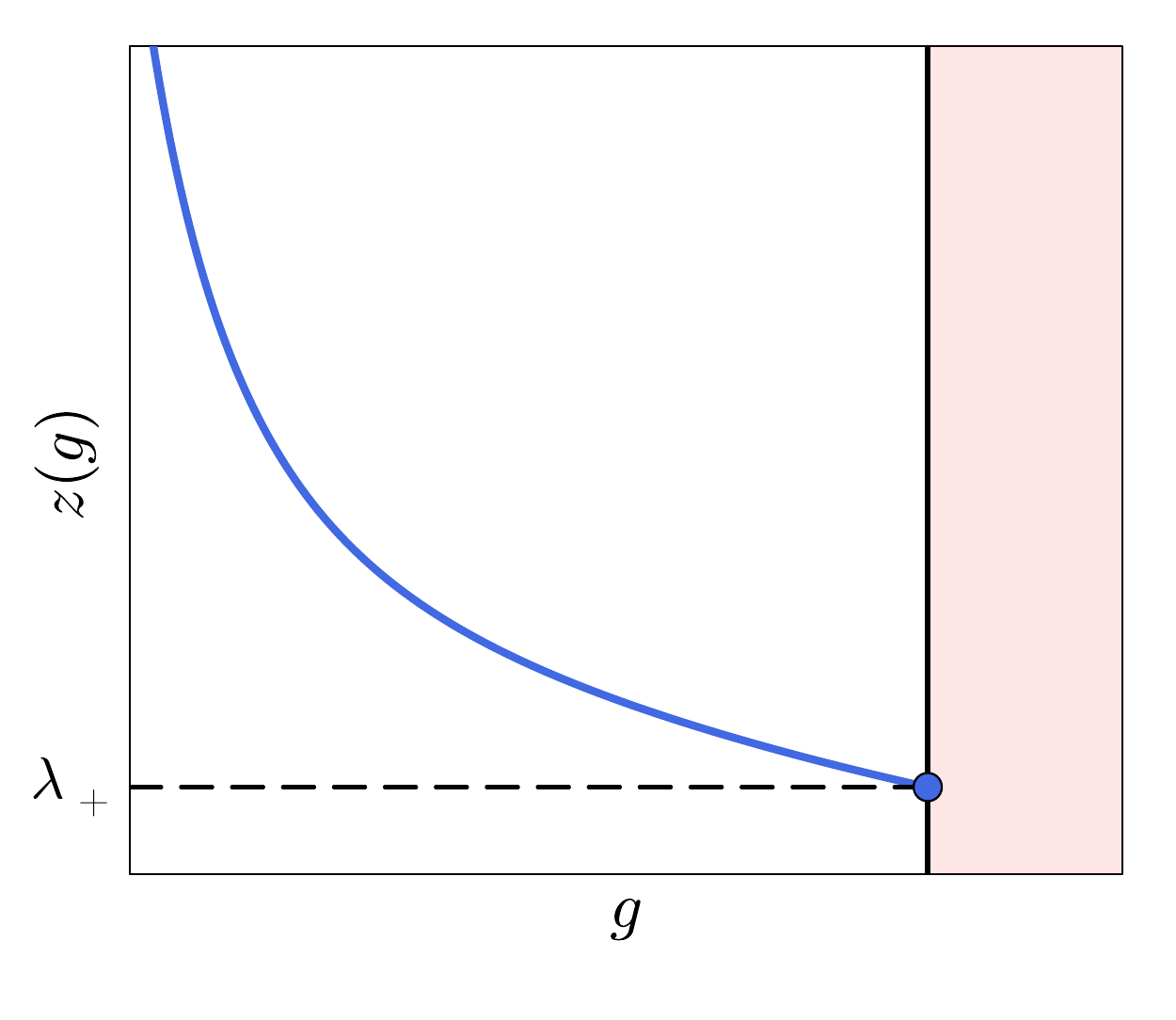}
    \caption*{\small Inverse resolvent $z(g)$}
\end{subfigure}
\hfill
\begin{subfigure}{0.32\textwidth}
    \centering
    \includegraphics[width=\linewidth]{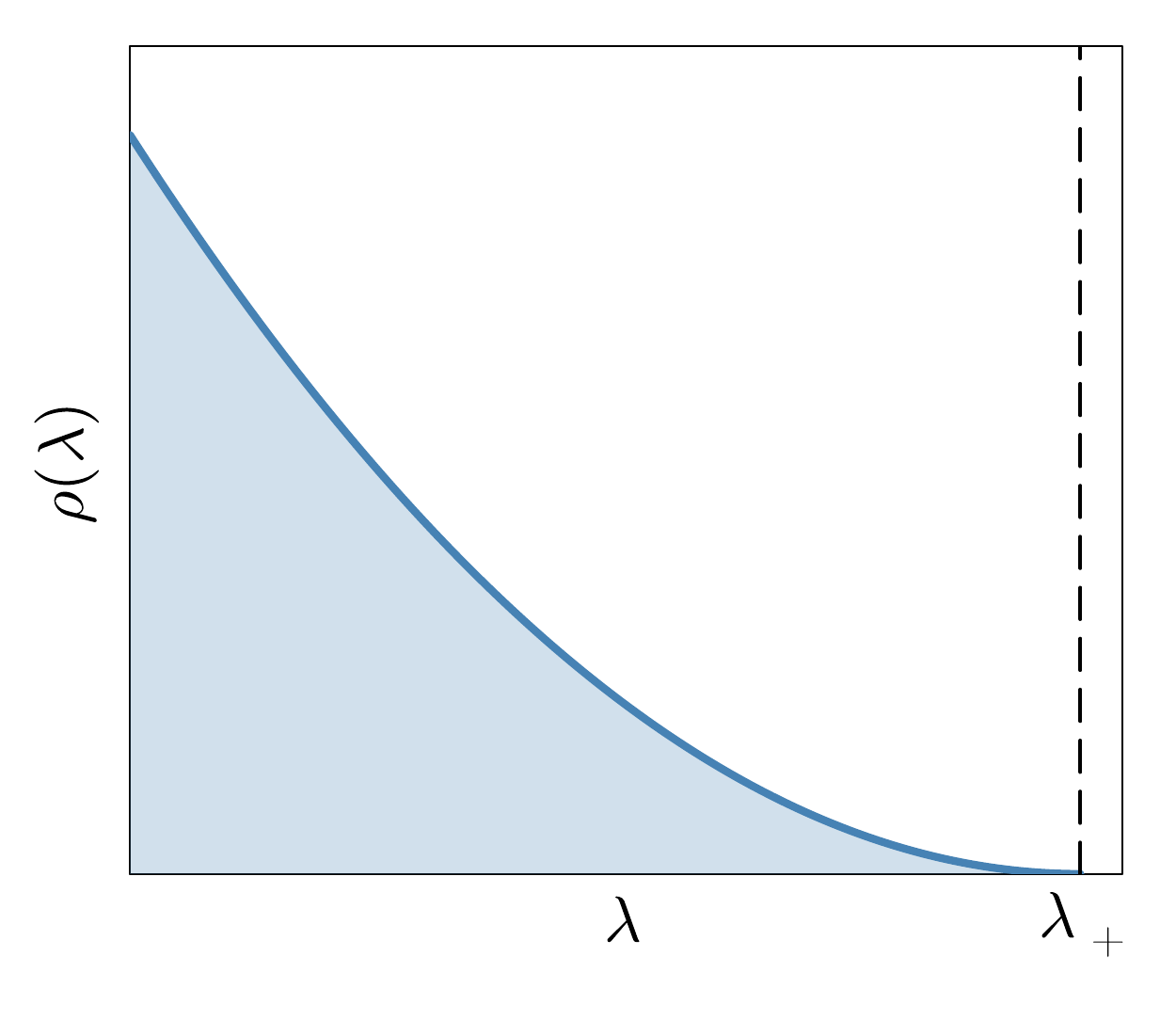}
    \caption*{\small Density $\rho(\lambda)$}
\end{subfigure}
\hfill
\begin{subfigure}{0.32\textwidth}
    \centering
    \includegraphics[width=\linewidth]{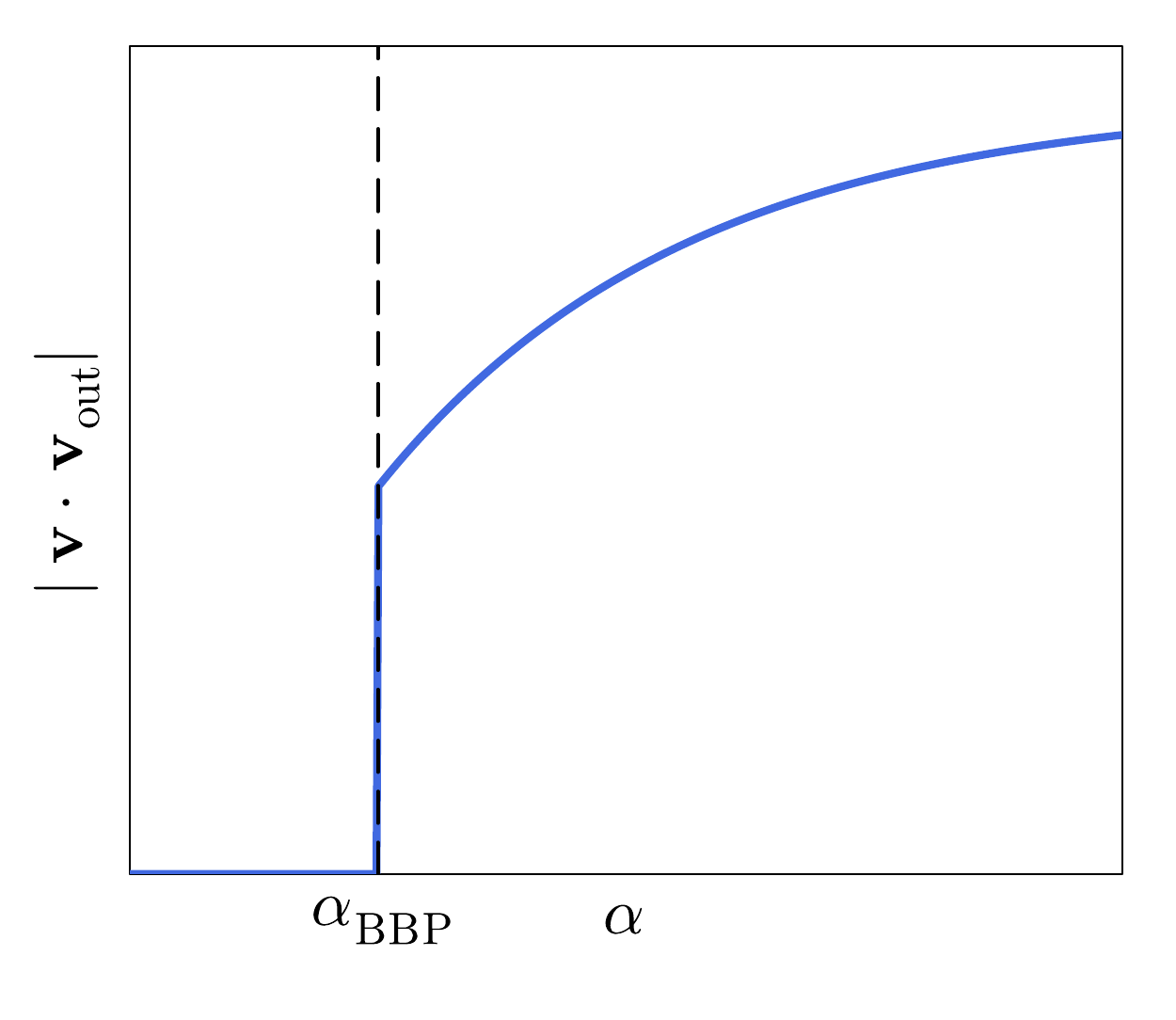}
    \caption*{\small Overlap $|\mathbf{v}\!\cdot\!\mathbf{v}_{\mathrm{out}}|$}
\end{subfigure}
\end{tcolorbox}

\caption{
\textbf{Continuous vs. Discontinuous BBP transitions}.  
Each row shows (left to right) the inverse resolvent map $z(g)$, the spectral density near the right edge, and the eigenvector overlap as a function of the spike strength. In the \textit{continuous} case, $z(g)$ develops a turning point ($dz/dg = 0$) at the edge, beyond which lies the incorrect branch, the density vanishes sub-linearly, and the overlap turns on smoothly. In the \textit{discontinuous} case, $z(g)$ reaches the edge without a turning point ($dz/dg \neq 0$), the density vanishes super-linearly at the edge, and the overlap exhibits a finite jump at onset.
}
\label{fig:bbp_main}
\end{figure*}

The spectrum of large random matrices is usually analyzed using the resolvent formalism \cite{potters2020first}. For a symmetric $N\times N$ matrix $\bm{A}$ with eigenvalues $\{\lambda_k\}$, the resolvent is defined as $(z\bm{I}-\bm{A})^{-1}$. The normalized trace of the resolvent is called the Stieltjes transform and is given by
\begin{equation}
    g_N(z) = \frac{1}{N}\sum_{k=1}^N \frac{1}{z-\lambda_k}.
\end{equation}
When $N$ goes to infinity, if the eigenvalues become dense and converge to a limiting distribution with density $\rho(\lambda)$ that only depends on the matrix ensemble of $\bm{A}$, the Stieltjes transform becomes
\begin{equation}
g(z)=\int_{\operatorname{supp}(\rho)}\frac{\rho(\lambda)}{z-\lambda}\,d\lambda,
\label{eq:stieltjesDef}
\end{equation}
which is well defined and real for all $z$ on the real axis, excluding points inside the support of $\rho(\lambda)$. In this case, the resolvent encodes all relevant properties of the eigenvalue distribution for the ensemble of $\bm{A}$. In particular, the spectral density $\rho(\lambda)$ can be recovered from the Stieltjes transform through the Stieltjes inversion formula
\begin{equation}
\rho(\lambda)=\lim_{\epsilon\to0^+}\frac{1}{\pi}\,\mathrm{Im}\,g(\lambda-i\epsilon).
\label{eq:genericspectrum}
\end{equation}

For a given random matrix ensemble, the Stieltjes transform $g(z)$ is usually determined in the large-$N$ limit by a self-consistent equation of the form \cite{potters2020first}:
\begin{equation}
F[g(z),z]=0.
\label{eq:generic_selfConsistent}
\end{equation}
In the following, we assume that the continuous spectrum occupies a single interval $[\lambda_-,\lambda_+]$. 
If the support consists of several disjoint components, the analysis applies separately on each of them.

\subsection{Structure of Spectral Edges}
\label{sec:spectralEdges}

To find the spectral edges $\lambda_\pm$, one looks for the smallest and largest real values of $z$ for which Eq.~\eqref{eq:generic_selfConsistent} ceases to admit a real solution.  There are two distinct mechanisms through which this may happen. In the following, we focus on the right edge $\lambda_{+}$, but the same reasoning can also be applied to the left edge $\lambda_-$. \\

The first corresponds to a breakdown of the implicit function theorem. Indeed, as long as $\partial_g F(g,z) \neq 0$, the theorem guarantees a locally unique and smooth solution $g(z)$. When $\partial_g F(g,z) = 0$, this condition fails and the solution branch develops a vertical tangent in the $(z,g)$ plane. Equivalently,
\begin{equation}
 \left| g'(z) \right|
 =
 \left| - \frac{\partial_z F(g,z)}{\partial_g F(g,z)} \right|
 \to \infty ,
\end{equation}
indicating the coalescence of two real solution branches. Since $g(z)$ is single-valued, there exists only one correct branch compatible with Eq.~\eqref{eq:stieltjesDef}, which is uniquely identified by the asymptotic behavior $g(z) \sim 1/z$ as $z \to \infty$. In terms of the inverse function $z(g)$, this implies that the correct branch corresponds to the region to the left of the minimum of $z(g)$ (where $z'(g) = 0$), while the branch to the right corresponds to an incorrect solution. We now show that this mechanism corresponds to the case in which the spectral density vanishes sub-linearly at the edge, i.e.,
\begin{equation}
\rho(\lambda)\sim (\lambda_+-\lambda)^{a-1}, \qquad 0  <a \leq 2,
\end{equation}
including the classical square-root case $a=3/2$. This is sketched in the first row of Figure \ref{fig:bbp_main}. The link between the exponent $a$ and the implicit-function breakdown follows from the edge asymptotics. 
At the spectral edge $\lambda_+$, we evaluate $g'(\lambda_+)$ directly using the definition of $g(z)$
\begin{equation}
g'(z)= - \int_{\lambda_-}^{\lambda_+}\frac{\rho(\lambda)}{(z-\lambda)^2}\,d\lambda,
\end{equation}
and the power-law behavior of the density:
\begin{equation}
\rho(\lambda)\sim(\lambda_+-\lambda)^{a-1} \quad \text{for } \lambda\to\lambda_+.
\end{equation}
Setting $x=\lambda_+-\lambda$, we get
\begin{equation}
g'(\lambda_+)
\sim
-\int_{0}^{\lambda_+ - \lambda_-} \frac{x^{\,a-1}}{x^2}\,dx
=
-\int_{0}^{\lambda_+ - \lambda_-} x^{\,a-3}\,dx.
\end{equation}
For $a\leq 2$ this integral diverges in the region $x \sim 0$, implying $g'(\lambda_+)\to-\infty$ and a turning point $z'(g)=0$.

The second mechanism arises when the self-consistent equation itself ceases to be well defined (for instance, because $F(g,z)$ becomes singular). This occurs when the density vanishes faster than linearly at the edge,
\begin{equation}
\rho(\lambda)\sim (\lambda_+-\lambda)^{a-1}, \qquad a>2,
\end{equation}
so that the above integral converges, $g'(z)$ remains finite, and $z'(g)\neq 0$ at the edge. In this case, the resolvent stays monotonic and the spectrum terminates smoothly without branch merging. In Figure \ref{fig:bbp_main}, this situation is sketched in the lower panel.

In practice, the spectral edge is found either when $\partial_gF(g,z)=0$ (equivalently $z'(g)=0$), or when the solution reaches the boundary of the domain where $F$ is defined, whichever comes first. The way this happens distinguishes sublinear from superlinear edges and directly dictates whether the BBP transition is continuous or discontinuous, as we will see in the next section.

\subsection{BBP Transition: General Theory}

The Baik–Ben Arous–Péché (BBP) transition \cite{baik2005phase} occurs when a finite-rank perturbation is added to a large random matrix and produces an \emph{outlier} eigenvalue separating from the spectral bulk. At the same time, the associated eigenvector acquires a nonzero overlap with the perturbation direction, indicating that a signal becomes detectable against the noise.
Consider the spiked model
\begin{equation}
    \bm{M}_\alpha = \bm{A} + \alpha\, \bm{v}\bm{v}^\top ,
\end{equation}
where $\bm{v}$ is a vector independent of $\bm{A}$ and with unit norm while $\alpha$ controls the spike strength.  
An outlier eigenvalue $\lambda_{\rm out}$, when it exists, satisfies \cite{potters2020first}:
\begin{equation}
    g(\lambda_{\rm out}) = 1/\alpha,
    \label{eq:bbp_condition}
\end{equation}
where $g(z)$ is the Stieltjes transform of the unperturbed ensemble $\bm{A}$. 
The corresponding eigenvector overlap is
\begin{equation}
    |\bm{v}\!\cdot\!\bm{v}_{\rm out}|^2
    = -\frac{1}{\alpha^2 g'(\lambda_{\rm out})} .
    \label{eq:overlap}
\end{equation}

The BBP critical point $\alpha_{\rm BBP}$ is defined as the smallest spike strength for which an outlier appears, i.e.
\begin{equation}
\lambda_{\rm out}(\alpha_{\rm BBP}) = \lambda_+ .
\end{equation}

Since the outlier collides with the edge at $\lambda_+$, the behavior of $g'(z)$ at the edge dictates the nature of the transition. From the analysis above,
\[
|g'(\lambda_+)| =
\begin{cases}
\infty, & a \leq 2 \quad\text{(sublinear edge)}, \\
\text{finite}, & a > 2 \quad\text{(superlinear edge)}.
\end{cases}
\]
Therefore, we have that:
\begin{itemize}
    \item for $a \le 2$, $|g'(z)|\to\infty$ at the edge, so the overlap in Eq.~\eqref{eq:overlap} vanishes at the transition: the BBP transition is \emph{continuous};
    \item for $a>2$, $|g'(\lambda_+)|$ remains finite, and the overlap jumps to a nonzero value at $\alpha_{\rm BBP}$: the BBP transition is \emph{discontinuous}.
\end{itemize}
In Figure \ref{fig:bbp_main}, these aspects are summarized for a generic ensemble.

\section{The deformed GOE ensemble}

\begin{figure*}
    \centering
    \includegraphics[width=0.7\linewidth]{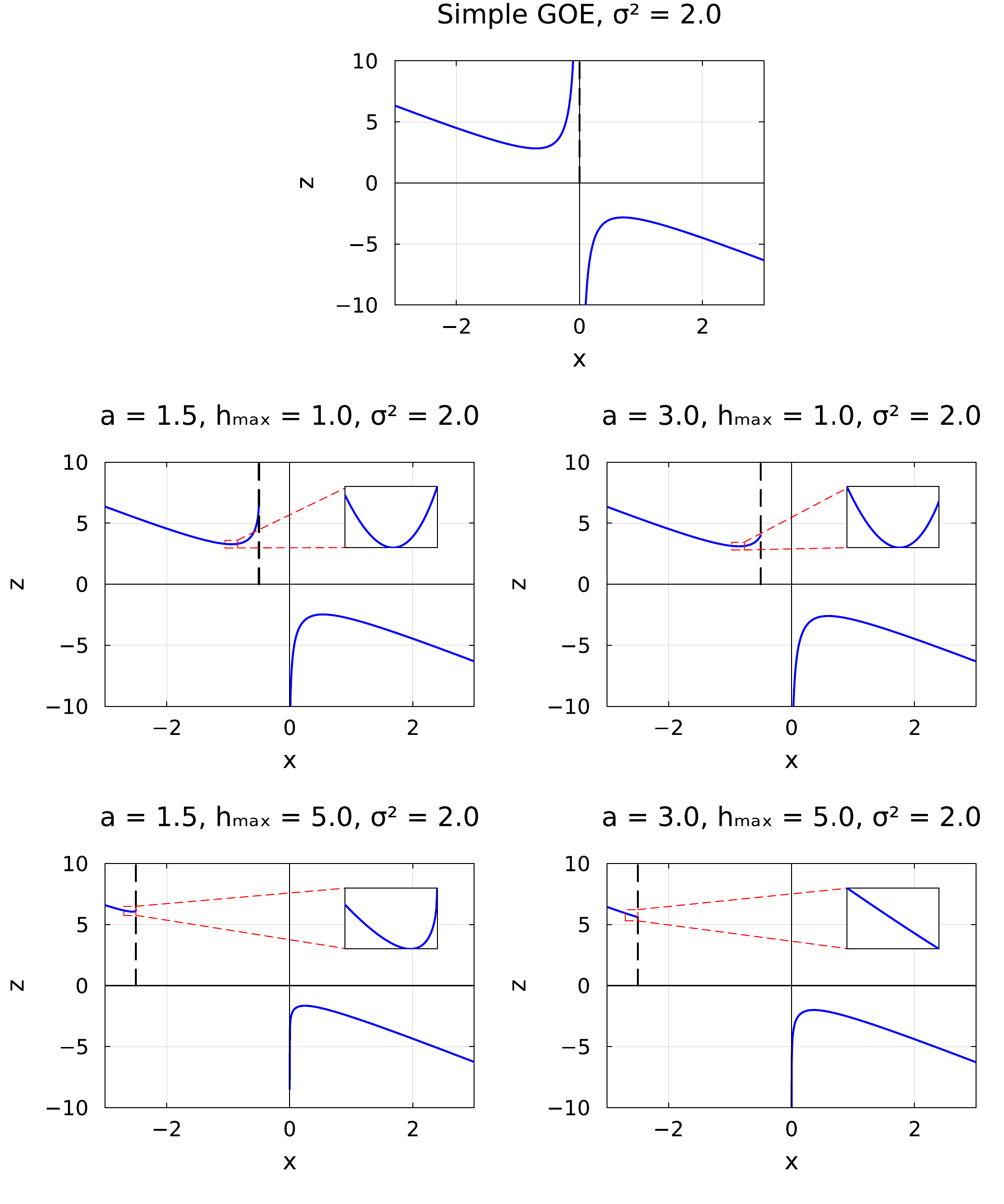}
    \caption{\textbf{The origin of different tails of spectral densities in the Deformed GOE ensemble.} 
    The function $z(x)$ is plotted against the auxiliary variable $x = g - z/\sigma^2$ for different values of $a$ and  $h_{\max}$. In the first line, the plot for a standard GOE ($h_{max} = 0$) is shown. \textit{Left column ($a < 2$):} The function $z(x)$ always exhibits a turning point ($z'(x)=0$) and the incorrect branch diverges at the domain boundary $x_c = -h_{\max}/\sigma^2$, implying a standard square-root edge in all three cases. \textit{Right column ($a > 2$):} The derivative $z'(x)$ is finite at the domain boundary, and depending on the values of $h_{max}$ and $\sigma$ can be positive (mid panel), leading again to a square-root edge, or negative (bottom panel), leading to a super-linear edge.}
    \label{fig:GOEnonGOEtail}
\end{figure*}

We first consider a model consisting of a Gaussian Orthogonal Ensemble (GOE) matrix perturbed by a diagonal matrix $\bm{D}$ and a rank-one spike. This setup naturally extends previous models explored in the literature~\cite{lee2013local,lee2016extremal,krajenbrink2021tilted,bouchbinder2021low,edwards1976eigenvalue}. The diagonal entries $h_i$ of $\bm{D}$ are drawn from a distribution $P_D(x)$, characterized primarily by its power-law behavior near the upper edge: $P_D(x)\sim (h_{\rm max}-x)^{a-1}$ as $x \to h_{\rm max}$. For numerical applications, we will adopt the specific form
\begin{equation}
    P_D(x) = a(h_{\rm max}-x)^{a-1}h_{\rm max}^{-a}, \quad \text{for } 0\le x\le h_{\rm max}.
    \label{eq:diagonal_distribution}
\end{equation}
The resulting random matrix $\bm{M}$ is defined by
\begin{equation}
    M_{ij} = h_{i}\delta_{ij} + \sigma W_{ij} + \alpha v_i v_j,
    \label{eq:deformedGOE_ensemble}
\end{equation}
where $\bm{v}$ is a random vector of unit norm and $\bm{W}$ is a symmetric matrix with Gaussian entries $W_{ij}\sim \mathcal{N}(0,1/N)$. The parameters $\sigma$ and $\alpha$ control the strength of the Gaussian noise and the rank-one perturbation, respectively, relative to the diagonal disorder. For $\alpha=0$, this class of random matrices has been rigorously studied in Refs.~\cite{lee2013local, lee2016extremal} and within the physics literature in Refs.~\cite{krajenbrink2021tilted, bouchbinder2021low, franz2022delocalization}. In contrast, without the diagonal deformation (i.e., $h_i=0$ for all $i$), the matrix behaves as a standard model for the BBP transition, where an isolated eigenvalue separates from the right edge of the Wigner semicircle for sufficiently large $\alpha$~\cite{edwards1976eigenvalue}. In the following, we analyze the BBP transition for the ensemble defined in Eq.~\eqref{eq:deformedGOE_ensemble} using the general framework developed in Sec.~\ref{sec:spectral}. The key insight is that the statistics of the diagonal disorder $h_i$ control the nature of the spectral edge, thereby determining whether the BBP transition is continuous or discontinuous.

It can be shown, using for instance free probability techniques (see Appendix~\ref{sec:app_deformedGOE_derivation}), that the limiting Stieltjes transform of the deformed GOE ensemble $g(z)$ satisfies the self-consistent equation:
\begin{equation}
    g(z)
    - \int_0^{h_{\text{max}}} dh\,P_D(h)\,\frac{1}{z - h - \sigma^2 g(z)} =0.
    \label{eq:self_deformedGOE}
\end{equation}
A useful way to study and characterize the solutions is to plot 
$z$ as a function of $x=g-z/\sigma^2$. The relation between $x$ and $z$ can be obtained directly from Eq.~\eqref{eq:self_deformedGOE}
\begin{equation}
    z(x)=-\sigma^2\int_0^{h_{\text{max}}} dh P_D(h)\left[ x+\frac{1}{h+\sigma^2 x} \right].
    \label{eq:selfz_deformedGOE}
\end{equation}
The self–consistent equation \eqref{eq:selfz_deformedGOE} implies a constraint on the auxiliary
variable \(x = g - z/\sigma^2\): since the integrand must remain finite, the quantity
\((h + \sigma^2 x)\) in the denominator cannot vanish. Since \(h \in [0,h_{\max}]\), this yields
the domain of definition
\begin{equation}
     x < -\frac{h_{\max}}{\sigma^2}\; \cup\; x>0\,. 
    \label{eq:x_domain}
\end{equation}
Thus, $z(x)$ -- and consequently $z(g)$ -- is only defined for $x$ in these regions. We plot in the left and in the right column of Figure \ref{fig:GOEnonGOEtail} the curve $z(x)$ for the case of sublinear edge ($a<2$) and superlinear edge ($a>2$), respectively, for different values of $h_{\rm max}$. The case $h_{\rm max}=0$ (top panel) corresponds to a standard GOE matrix, as in that case all the elements of the diagonal matrix are zero. In order to determine which branch of the curve $z(x)$ is the correct one, i.e.\ it corresponds to the true $g(z)$, we recall that for large arguments the function $g(z)$ vanishes as $1/z$, which implies that  $z(x)\simeq-x\sigma^2$ for $z\rightarrow \pm \infty$. Therefore by starting from very large $z$ (or very negative $x$) one follows the branch on the left (or right) until the minimum (or maximum) of $z(x)$ 
is reached. 
The behavior of $\rho(\lambda)$ at the edge of the spectrum is completely determined by the behavior of $z(x)$ when $z$ approaches the edge from outside \cite{bouchbinder2021low}.

\paragraph{Wigner square-root edge: $dz/dx=0$ at the minimum.} This is the case of the usual GOE, where $dz/dx=0$ for $z$ at the edge of the spectrum (see top panel in Fig.~\ref{fig:GOEnonGOEtail}), the density of eigenvalues vanishes with a square root singularity. For $h_{\rm max}$ different from zero, i.e. the deformed matrix $M$, if $a<2$, as shown in the left panels of Fig.~\ref{fig:GOEnonGOEtail}, the situation remains similar to the previous one except that the derivative of the incorrect branch diverges at $x=-h_{\rm max}/\sigma^2$ for negative $x$: it can be shown~\cite{bouchbinder2021low} that its derivative $z'(x)$ is always positive and infinite, when evaluated at $x=-h_{\rm max}/\sigma^2$. Since the correct branch of $z(x)$ goes as $-x\sigma^2$ at very negative $x$ (with negative derivative $z'(x)\simeq-\sigma^2$), one also finds there is an intermediate $x$ where $dz/dx=0$. 
This is also what happens for the middle-right panel, except that $z'(x)$ is positive but finite at $x=-h_{\rm max}/\sigma^2$. In all these cases, the $z$ at which $dz/dx=0$ corresponds to the edge of the spectrum where the density of eigenvalues vanishes with a square root singularity, for all values of $h_{\rm max}$.

\begin{figure*}
    \centering
    \includegraphics[width=0.7\linewidth]{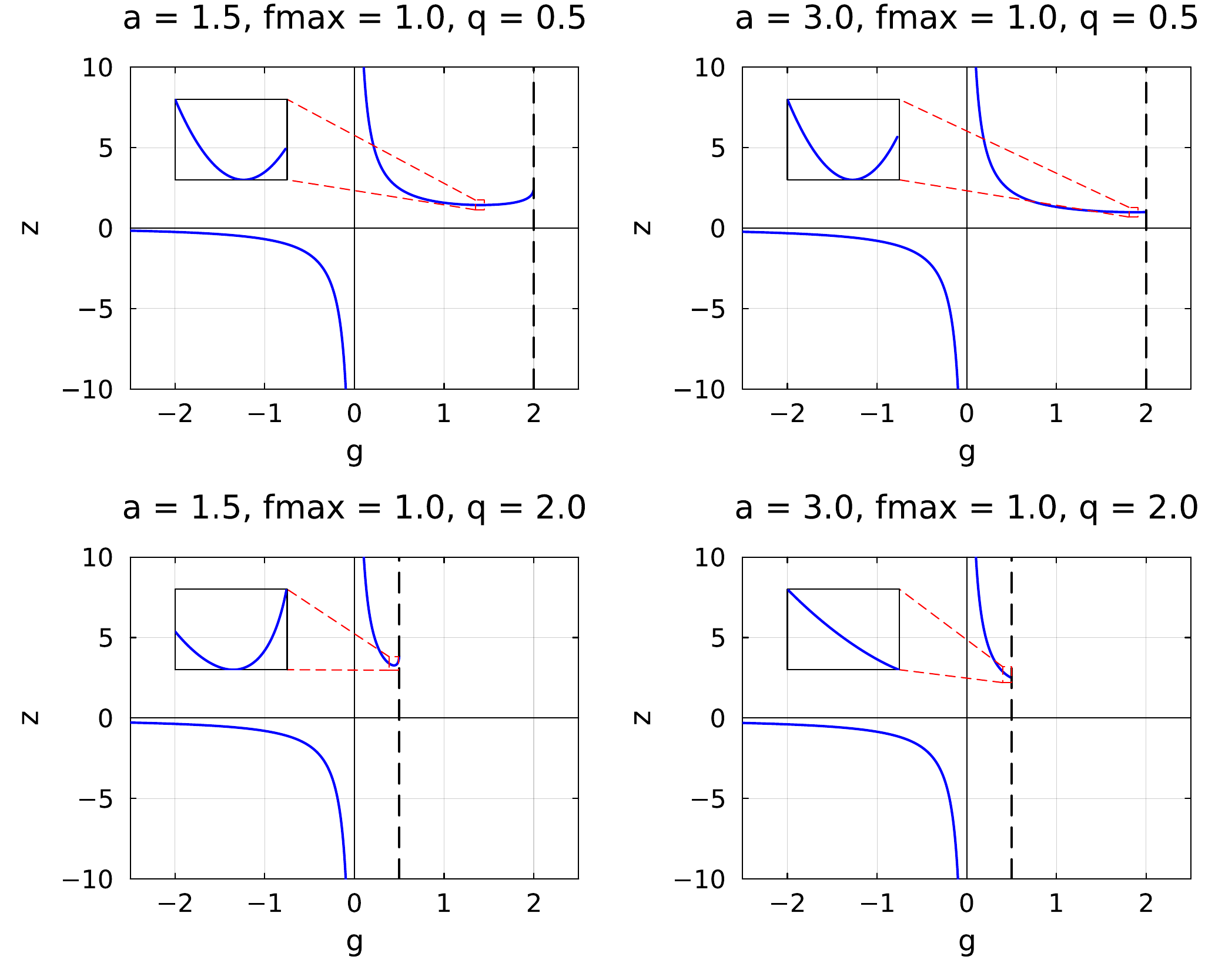}
    \caption{\textbf{The origin of different tails of spectral densities in the Reweighted Wishart ensemble.}
    The function $z(g)$ is plotted for different values of $a$ and the aspect ratio $q=N/T$. \textit{Left column ($a < 2$):} The branch with incorrect positive derivative $z'(g)$, diverging at the domain boundary $g_{\max} = (q f_{\max})^{-1}$, ensures the existence of a turning point with $z'(g)=0$ and consequently a standard square-root edge. \textit{Right column ($a > 2$):} The derivative $z'(g)$ is finite at the domain boundary, and depending on the value of $q$ can be positive (top panel), allowing for a turning point and a square-root edge, or negative (bottom panel), leading to a super-linear edge.}
    \label{fig:WishartNonWishartTail}
\end{figure*}

\paragraph{Super-linear edge: $dz/dx\ne 0$ at the minimum} In this case, the behavior at the edge is the same as the distribution of the diagonal elements of $D$, i.e., a power-law with exponent $a-1>1$ \cite{bouchbinder2021low}. This case corresponds to large enough diagonal deformations, i.e., small $\sigma$ or large $h_{\rm max}$ (see Appendix~\ref{sec:app_deformedGOE_condition} for details). One finds that $z'(x)$ is also negative at $x=-h_{\rm max}/\sigma^2$ and $z(x)$ is a monotonic decreasing function. In this case, the minimum of $z(x)$ is reached at $x=-h_{\rm max}/\sigma^2$ where $z'(x)$ is finite as shown in the lower-right panel of Fig. \ref{fig:GOEnonGOEtail}.

According to the general theory, when the spectral density vanishes sub-linearly at the edge, the BBP transition induced by the rank-one perturbation $\alpha \bm{v}\bm{v}^\top$ will be continuous, while if the edge is super-linear, the overlap between the outlier eigenvector and the signal will jump to a finite value at the transition.

\section{Reweighted Wishart ensemble}

We now turn to a second class of random matrices that can exhibit discontinuous BBP transitions: the \emph{reweighted Wishart ensemble}. These matrices arise naturally in high-dimensional statistics and signal processing, particularly in problems like phase retrieval and covariance estimation with heterogeneous measurements \cite{mondelli2018fundamental}.

The reweighted Wishart matrix $\bm{W}$ of size $N \times N$ is constructed from $T$ independent sample vectors $\bm{\xi}^\mu$ as
\begin{equation}
    W_{ij} = \frac{1}{T} \sum_{\mu=1}^{T} f_\mu \, \xi_i^\mu \xi_j^\mu + \alpha\, v_i v_j,
    \label{eq:reweighted_wishart}
\end{equation}
where the components $\xi_i^\mu \sim \mathcal{N}(0,1)$ are i.i.d. Gaussian variables, $v_i$ are the components of a random unit vector $\bm{v}$, and the weights $f_\mu$ are drawn from a distribution $P_f$ with bounded support $[f_{\min}, f_{\max}]$. In this context, $T$ represents the number of samples (or observations), and we consider the high-dimensional limit where $N, T \to \infty$ with a fixed aspect ratio $q = N/T$. Similarly to the deformed GOE case, the crucial feature governing the BBP transition is the behavior of $P_f$ near its maximum $f_{\max}$. We focus again on distributions with a power-law tail $P_f(f) \sim a (f_{\max} - f)^{a-1}$, as $f \to f_{\max}$.

Using techniques similar to the GOE case (see Appendix \ref{app:app_rewightedWishart_derivation} for details), one can obtain an equation for the inverse of the Stieltjes transform of the reweighted Wishart matrix $z(g)$:
\begin{equation}
z(g) = \mathbb{E}_f\left[\frac{f}{1 - q f g}\right] + \frac{1}{g},
\label{eq:zeta_function}
\end{equation}
where the expectation is over the weight distribution $P_f$. We assume for simplicity that $f_{min} = 0 < f_{max}$ and, in numerical implementations, we will use the same function form of Eq.~\eqref{eq:diagonal_distribution}:
\begin{equation}
    P_f(x) = a(f_{\rm max}-x)^{a-1}f_{\rm max}^{-a}, \quad \text{for } 0\le x\le f_{\rm max}.
\end{equation}

In this case, the function $z(g)$ is always well defined for $g \in \left(
-\infty\; ,0\right)\cup\left(0, g_{\max}\right)$, with
\begin{equation}
    g_{\max} = \frac{1}{q f_{\max}}.
\end{equation}
This derivation can be easily generalized to any distribution $P_f$, provided that it is bounded and has a power-law tail. We again distinguish two possible scenarios.

\paragraph{Square-root edge: $dz/dg=0$ at the minimum.}
If the function $z(g)$ has a local minimum at $\bar{g} \in (0, g_{\max})$, the spectrum has a standard square-root singularity at the edge $\lambda_{\max} = z(\bar{g})$. This minimum exists provided that $z'(g_{\max}) > 0$. As detailed in Appendix~\ref{app:app_rewightedWishart_condition}, this condition holds either for $a<2$, where the derivative of the incorrect branch of $z'(g)$ diverges at $g_{\max}$ (see left column of Fig.~\ref{fig:WishartNonWishartTail}), or for $a>2$ with $q$ sufficiently small, where the derivative $z'(g_{\max})$ is finite and positive (see top-right panel of Fig.~\ref{fig:WishartNonWishartTail}). In all these cases, the correct branch of $z(g)$ is identified by tracing the curve from the asymptote at $g=0$ (where $z \to \infty$) up to the minimum $\bar{g}$, beyond which the solution becomes incorrect.

\paragraph{Super-linear edge: $dz/dg \neq 0$ at the minimum.} When this happens, the upper band edge is not a square root, but rather inherits the power-law behavior of the distribution $P_f$ of the weights. In this case, the right edge of the spectrum is simply located at
\begin{equation}
\lambda_{+} = z(g_{\max}) = f_{\max} \left[q + \mathbb{E}_f\left(\frac{f}{f_{\max} - f}\right)\right]\,,
\end{equation}
and, when $P_f(f) \sim (f_{\max} - f)^{a-1}$, the eigenvalues distribution tail will correspond to 
\begin{equation}
\rho(\lambda) \sim (\lambda_{+} - \lambda)^{a-1}\,.
\end{equation}
Also in this case, the BBP transition in this ensemble will be continuous in the case of a sub-linear edge ($a\leq 2$) and discontinuous when the edge is super-linear ($a>2$).

\section{Finite-Size Scaling Theory of Discontinuous BBP Transitions}

\begin{figure*}
    \centering
    \includegraphics[width=0.4\textwidth]{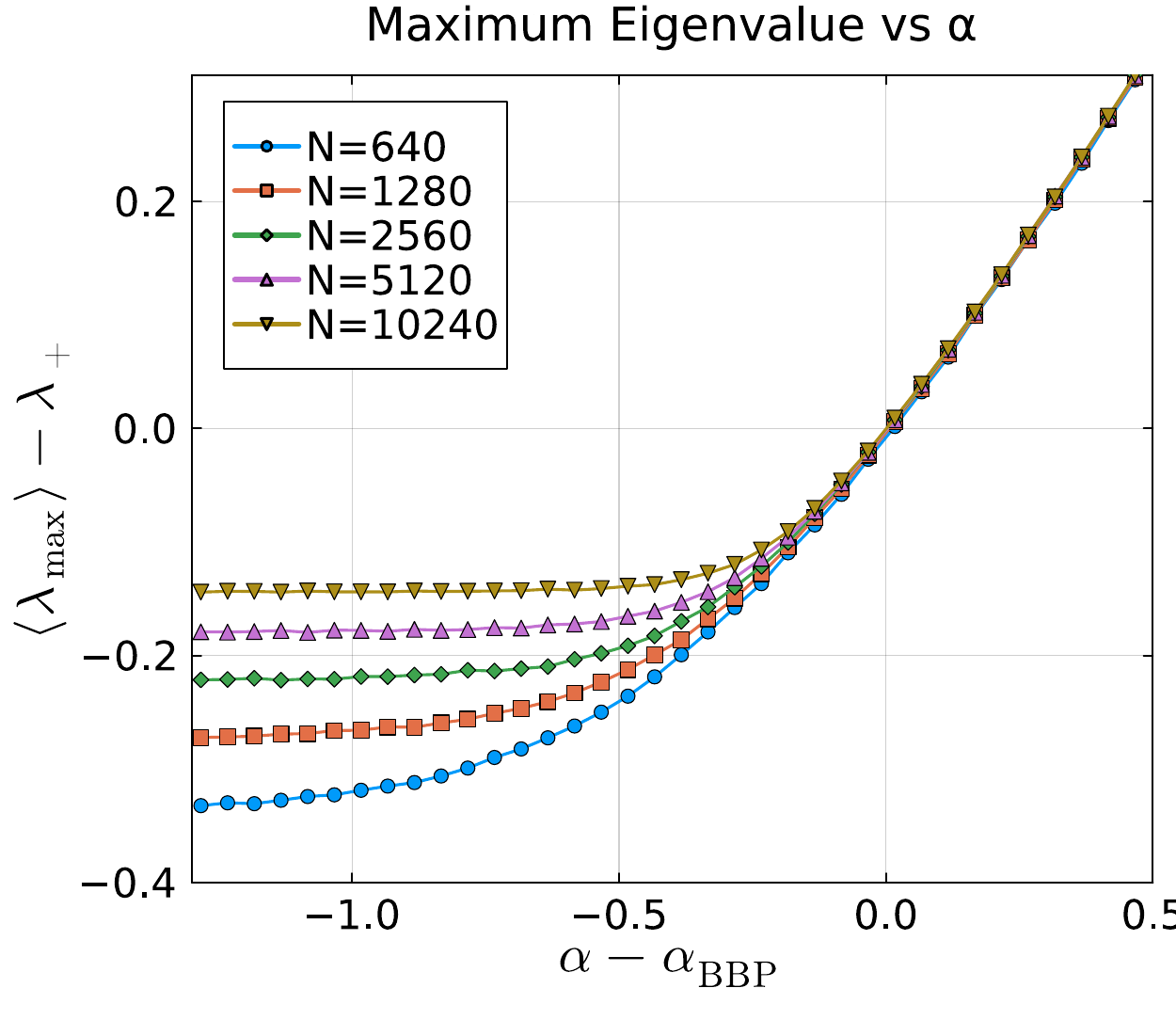}
    \hspace{1cm}
    \includegraphics[width=0.4\textwidth]{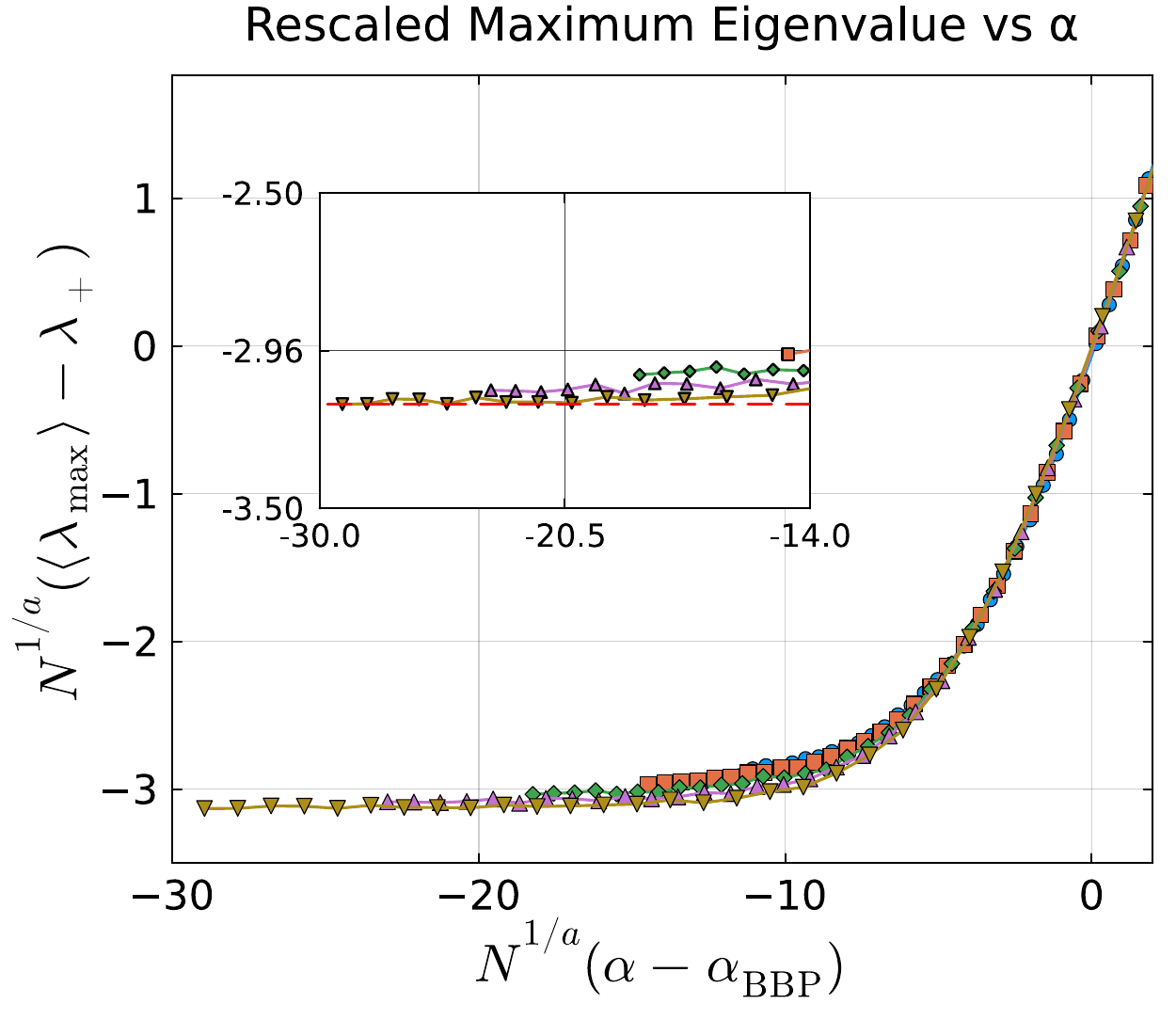}\\
    \includegraphics[width=0.4\textwidth]{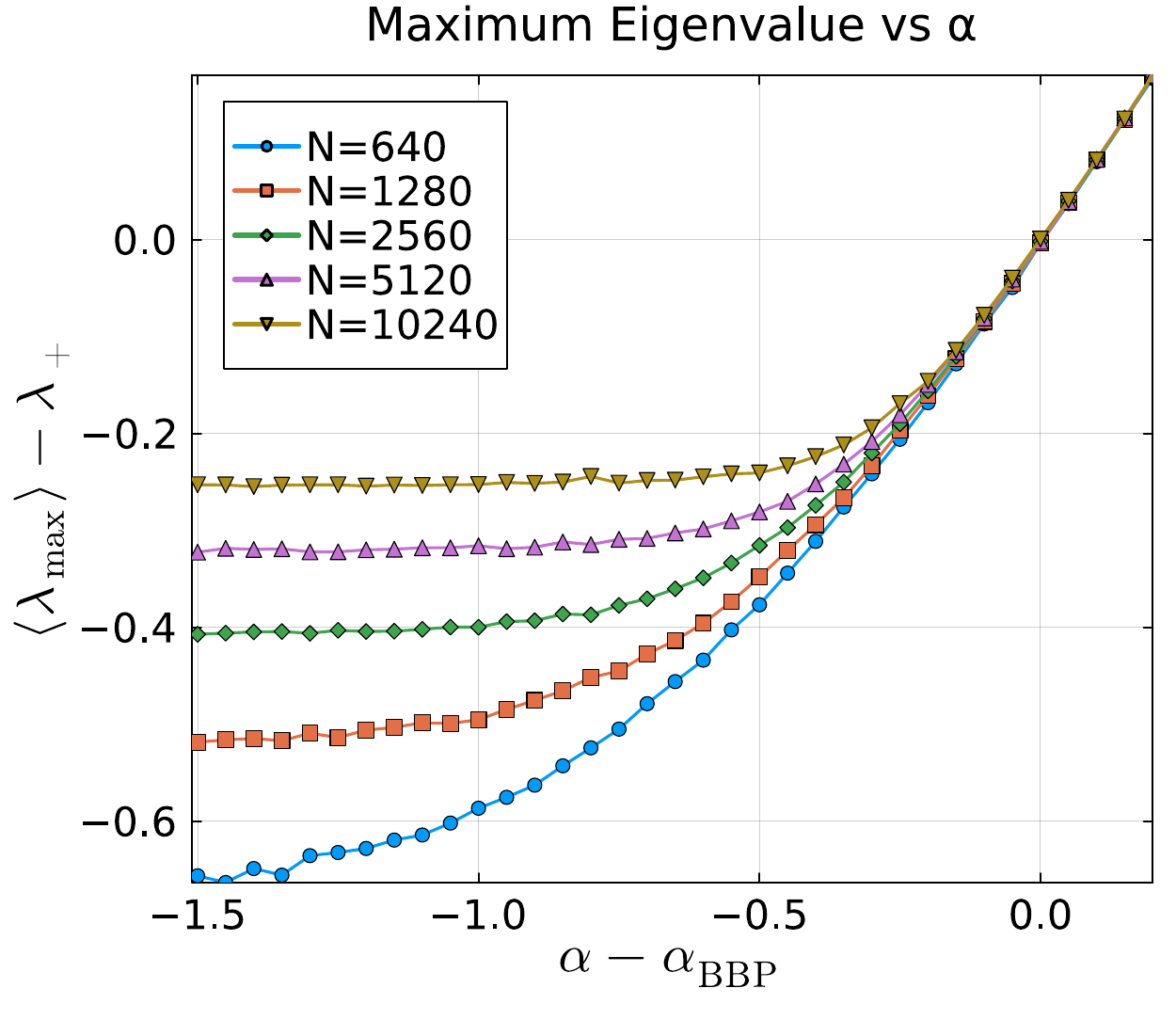}
    \hspace{1cm}
    \includegraphics[width=0.4\textwidth]{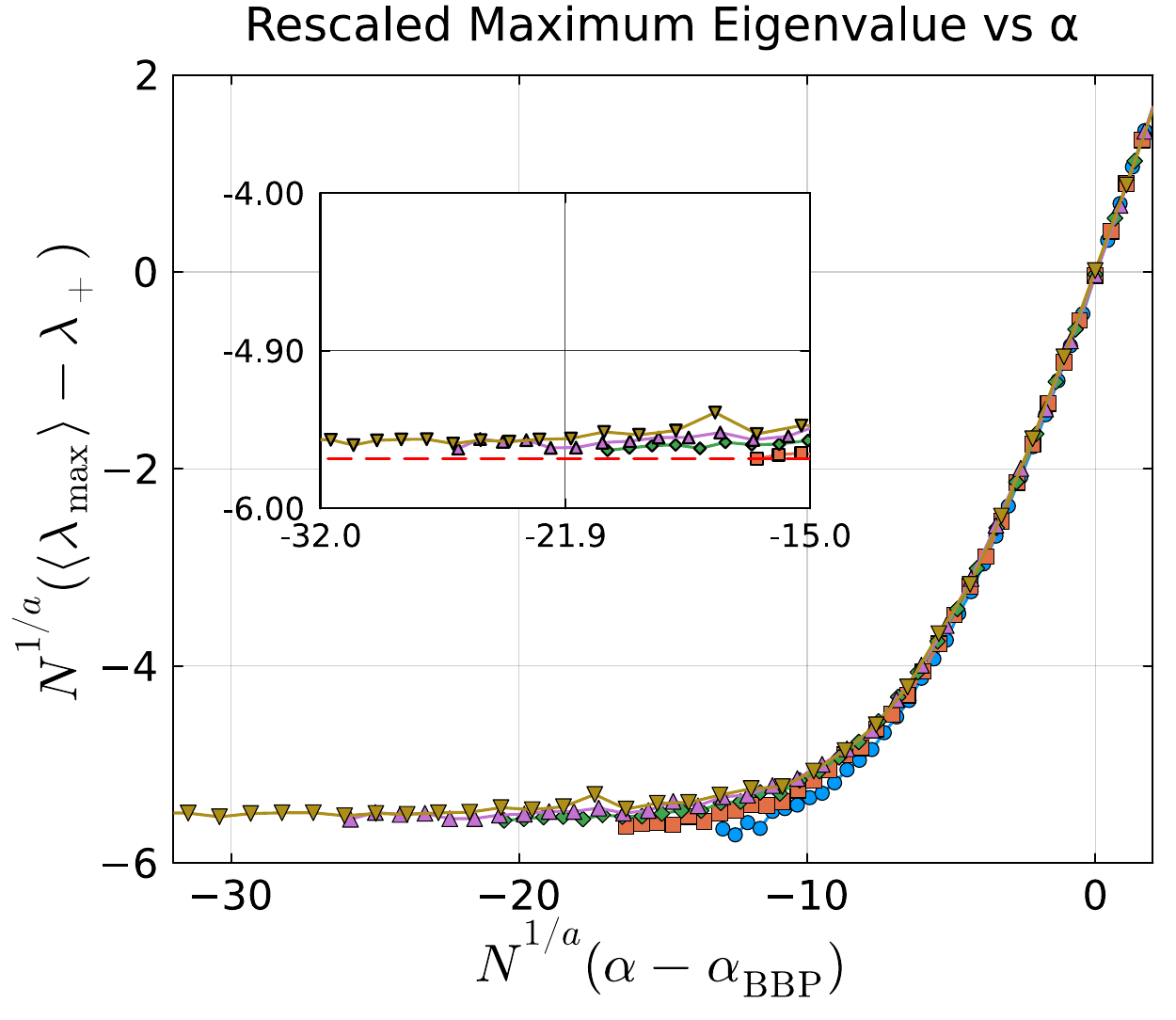}    
    \caption{\textbf{Finite-size scaling of the maximum eigenvalue.}
    \textit{Upper panels} are for the deformed GOE with parameters $a=3$, $h_{\max}=5$, and $\sigma^2 = 2$.
    \textit{Lower panels} are for the reweighted Wishart ensemble with parameter $a=3, f_{\max}=1$, and $q = 5$.
    In the \textit{left panels}, we plot the difference between the mean value of largest eigenvalue $\lambda_{\max}$ (averaged over $10^4$ samples) and the theoretical edge $\lambda_+$ computed in the large $N$ limit, as a function of the distance from the critical point, $\alpha - \alpha_{\mathrm{BBP}}$.
    In the \textit{right panels}, the same data is collapsed using the predicted scaling for both axes. The data collapse confirms that the fluctuations of the largest eigenvalue and the width of the critical region scale as $N^{-1/a}$, distinguishing this universality class from the standard Tracy-Widom $N^{-2/3}$ scaling. The insets report log-log plots of the left tail, showing the convergence to a constant (red dashed line).}
    \label{fig:scaling_collapse}
\end{figure*}

\begin{figure*}
    \centering
    \includegraphics[width=0.4\textwidth]{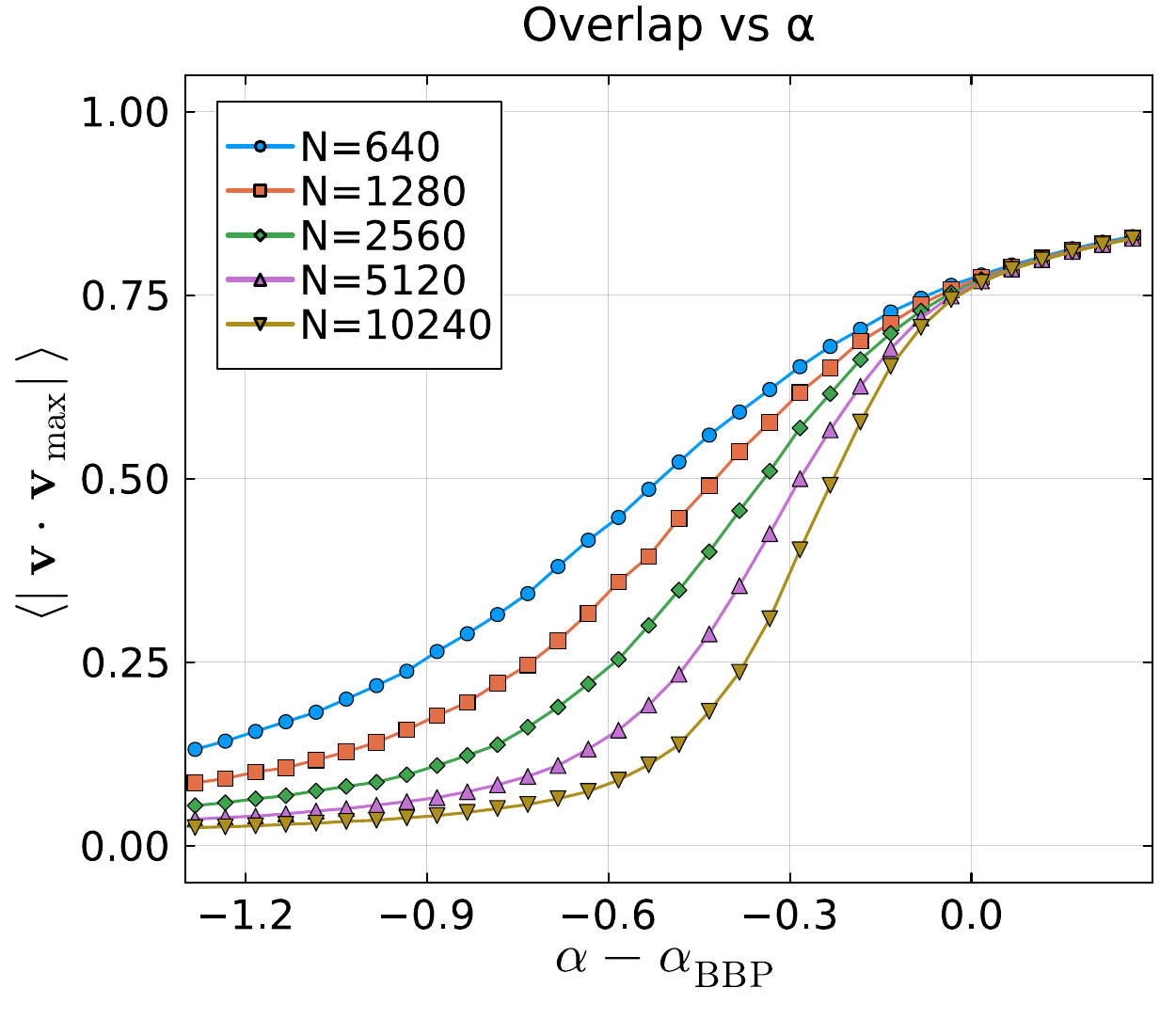}
    \hspace{1cm}
    \includegraphics[width=0.4\textwidth]{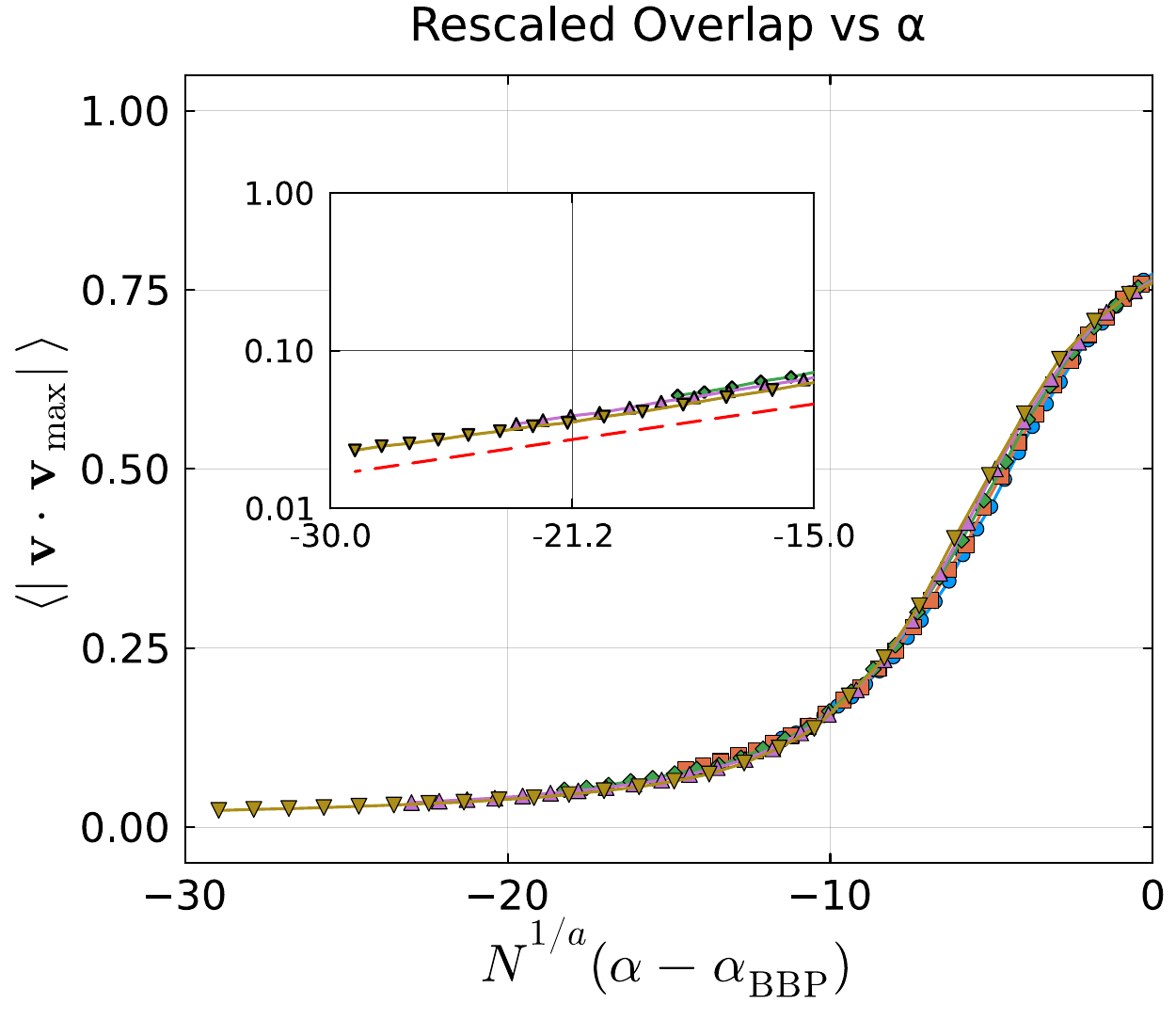}\\
    \includegraphics[width=0.4\textwidth]{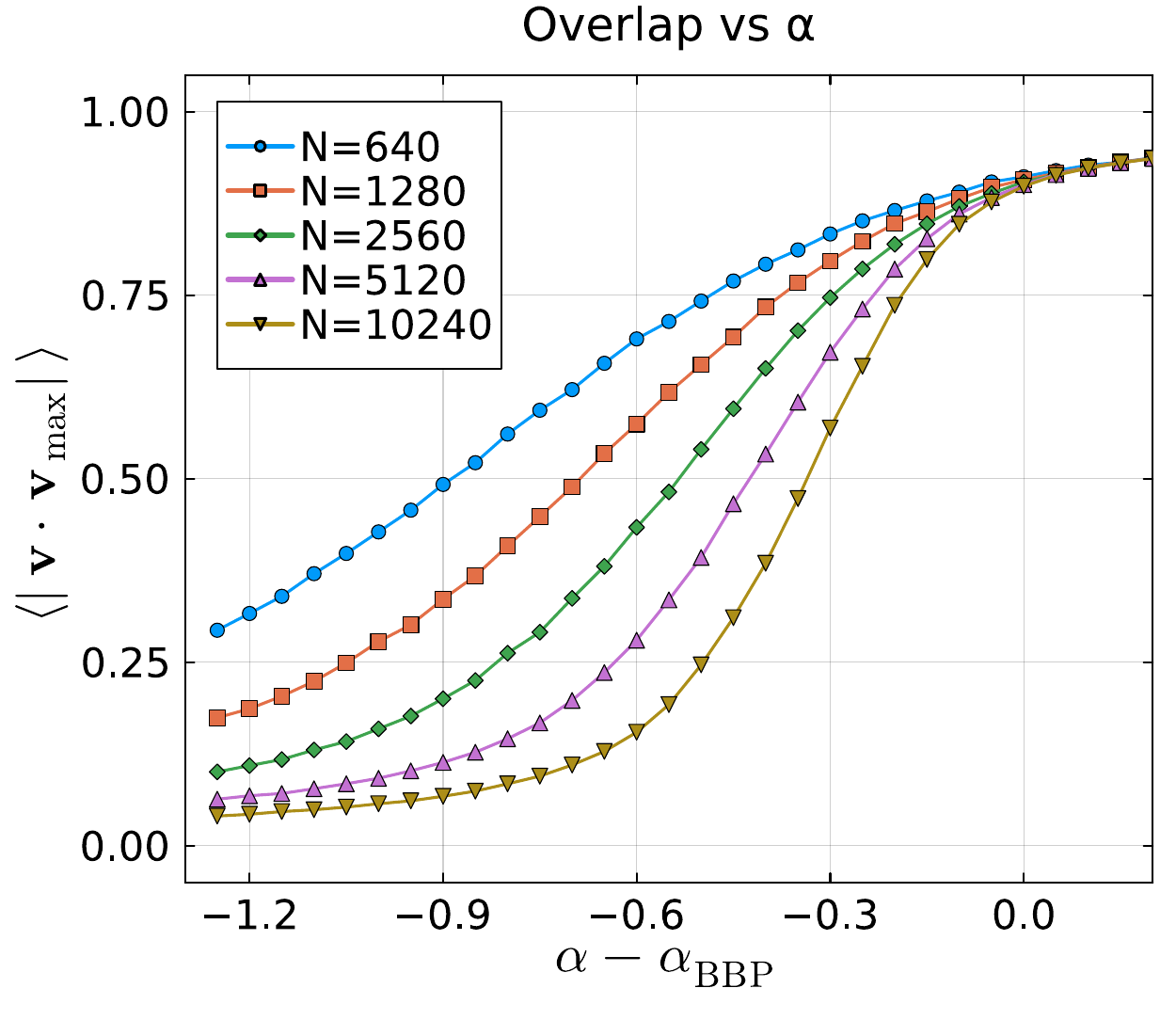}
    \hspace{1cm}
    \includegraphics[width=0.4\textwidth]{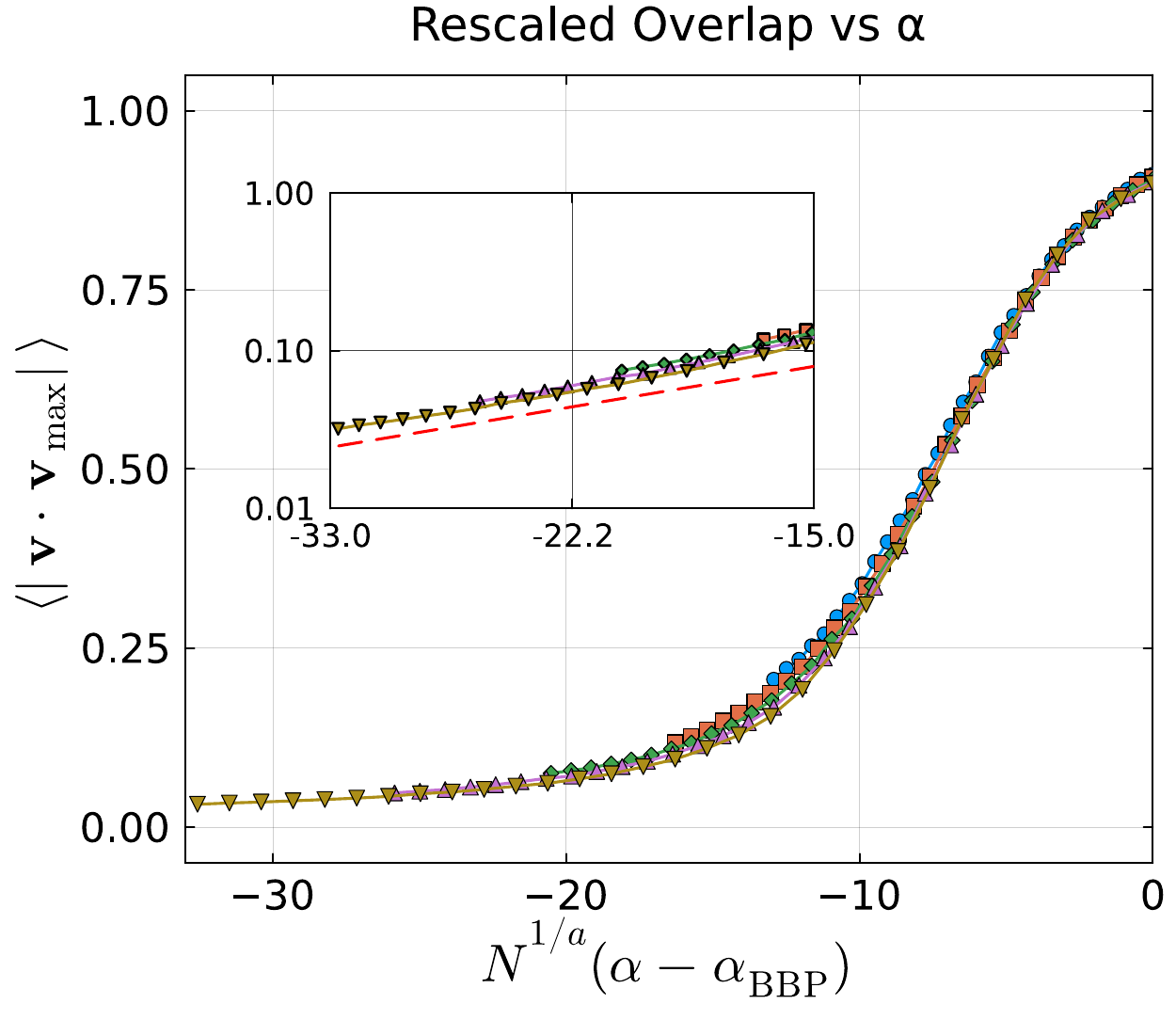}    
    \caption{\textbf{Finite-size scaling of the overlap.}
    \textit{Upper panels} are for the deformed GOE with parameters $a=3$, $h_{\max}=5$, and $\sigma^2 = 2$.
    \textit{Lower panels} are for the reweighted Wishart ensemble with parameter $a=3, f_{\max}=1$, and $q = 5$.
    In the \textit{left panels}, we plot the mean overlap between the principal eigenvector and the rank-one perturbation (i.e., the signal to be detected), as a function of the distance from the critical point, $\alpha - \alpha_{\mathrm{BBP}}$.
    In the \textit{right panels}, the same data is collapsed using the predicted scaling. The universal curve distinguishes the signal-dominated regime (linear behavior on the right) from the noise-dominated regime (left tail proportional to $|y|^{-a/2}$, see red dashed line in the log-log inset). The points are averages over $10^4$ samples.}
    \label{fig:overlap_collapse}
\end{figure*}

In the following, we study the role of finite-size effects in the discontinuous BBP transition. As we shall show, they are qualitatively different from the continuous case and can lead to significant modifications to the thresholds in realistic problems.

\subsubsection{Statistics of the largest eigenvalue for the matrix without the spike at finite size}

We start focusing on the role of finite-size effects on the largest eigenvalues for the matrix without a spike. We consider the case in which the density of eigenvalues has a power-law tail proportional to $(\lambda_+-\lambda)^{a-1}$ with $a>2$. For finite $N$, the largest eigenvalue $\lambda_1$ is not exactly equal to the value $\lambda_+$ derived in Sec.~\ref{sec:spectralEdges}, but it fluctuates at a certain distance to the left of it. The typical distance between $\lambda_1$ and the edge $\lambda_+$ is given by the following equation 
\[
\int_{\lambda_1}^{\lambda_+}\rho(\lambda)d\lambda = 1/N\,,
\]
which implies
\begin{equation}
\boxed{\lambda_+ - \lambda_1(N) \sim N^{-1/a}}
\label{eq:lambda1scaling}
\end{equation}
for a spectrum with a right tail $\rho(\lambda) \sim (\lambda_+-\lambda)^{a-1}$.

This simple argument can be made much more precise in the deformed GOE case $\bm{M} = \bm{D} + \sigma \bm{W}$. As proved in \cite{lee2016extremal}, the statistics of the first $k$ largest eigenvalues (with $k$ finite when $N\rightarrow \infty$) are given by the Weibull law: the statistics of the extremes of iid random variables with a bounded distribution. The reason is that the statistics of the largest eigenvalues of the matrix $\bm{D}$ is inherited by the matrix $\bm{M}$: the random part $\bm{W}$ is not enough to scramble it, as it produces fluctuations that are of the Tracy-Widom type, i.e. of the order of $N^{-2/3}$, which are much smaller than $N^{-1/a}$ \cite{lee2016extremal}.

\subsubsection{Statistics of the outlier eigenvalue at finite size}
\label{sec:outlierFSS}

Since in the discontinuous BBP case the derivative of the Stieltjes transform at the right edge is well defined and finite, we can start from Eq.~\eqref{eq:bbp_condition} to study how the position of the outlier $\lambda_\text{out}(\alpha)$ changes with $\alpha$ near the transition point $\alpha_{\mathrm{BBP}}$. Differentiating both sides of Eq.~\eqref{eq:bbp_condition} with respect to $\alpha$, we obtain
\begin{equation}
    \frac{d\lambda_{\mathrm{out}}}{d\alpha} \Big|_{\alpha = \alpha_{\mathrm{BBP}}} = -\frac{1}{\alpha_{\mathrm{BBP}}^2 g'(\lambda_+) } = m^2(\alpha_{\mathrm{BBP}})\,,
\end{equation}
where $m^2(\alpha_{\mathrm{BBP}})\equiv|\bm{v}\!\cdot\!\bm{v}_{\rm out}|^2$ is the overlap of the principal eigenvector with the signal (rank-one perturbation) at the BBP transition.
The non-zero value of $m^2(\alpha_{\mathrm{BBP}})$ implies that in the discontinuous case the eigenvalue separation scales linearly with $\Delta\alpha \equiv \alpha -\alpha_{\mathrm{BBP}}$ 
\begin{equation}
  \lambda_{\mathrm{out}}(\alpha) - \lambda_+ \simeq m^2(\alpha_{\mathrm{BBP}})\Delta\alpha +O(\Delta\alpha^2)\,.
  \label{eq:lambdaoutscaling}
\end{equation}

For finite $N$, since the maximum eigenvalue in the bulk $\lambda_1(N)$ is actually smaller than $\lambda_+$, we hypothesize that the linear behavior in Eq.~\eqref{eq:lambdaoutscaling} extends also below $\lambda_+$, in the region $\alpha < \alpha_{\mathrm{BBP}}$, where no outlier is present in the $N \to \infty$ limit (using the results of \cite{lee2016extremal}, this can be made precise for the GOE ensemble, as shown in Appendix \ref{sec:app_finiteN_GOE_outlier}).  
The extended linear behavior is expected to be the leading one in the interval $\lambda_1(N)<\lambda_\text{out}<\lambda_+$, that is on the scales $|\lambda_\text{out} - \lambda_+| \sim N^{-1/a}$ for large $N$, according to Eq.~\eqref{eq:lambda1scaling}.

As long as $\lambda_\text{out}>\lambda_1(N)$, the outlier is easily detectable. The limit of this region defines a size-dependent critical point $\alpha_c(N) \simeq \alpha_\text{BBP}-(\lambda_+-\lambda_1(N))/m^2(\alpha_{\mathrm{BBP}})$.
Combining it with Eq.~\eqref{eq:lambda1scaling} we get the following scaling
\begin{equation}
\boxed{\alpha_{\mathrm{BBP}} - \alpha_c(N) \sim N^{-1/a}}
\label{BBP_fss}
\end{equation}
Note that the value of $\alpha_c(N)$ also depends on the sample. The scaling above holds both for the average of $\alpha_c(N)$ and its fluctuations. 

\subsubsection{Statistics of the largest eigenvalue for the spiked matrix at finite size}

To establish a finite-size scaling (FSS) description for the maximum eigenvalue $\lambda_{\max}(\Delta \alpha, N)$ in the $\Delta\alpha < 0$ region, we first identify the appropriate scaling variables.
For $\Delta\alpha$ small enough, the maximum eigenvalue is the outlier and $\Delta \lambda_{\max} \equiv \lambda_{\max}(\Delta \alpha, N) - \lambda_+ \sim N^{-1/a}$, according the analysis in Sec.~\ref{sec:outlierFSS}.
Moreover, Eq.~\eqref{BBP_fss} implies $\Delta\alpha \sim N^{-1/a}$.

These arguments lead to the FSS ansatz for the average over the samples of $\Delta \lambda_{\max}$
\footnote{A more general FSS assumes the scaling 
$P(\lambda_{\max};\alpha,N) = H_1\left( N^{1/a}(\lambda_{\max}(\Delta \alpha, N)-\lambda_+),  N^{1/a}\Delta\alpha \right)N^{1/a}$ for the probability distribution of $\lambda_{\max}$, where $H_1(x,y)$ is a scaling function. This ansatz leads to the equation in the main text for $\langle \lambda_{\max} \rangle$, where $f(y)=\int dx\, H_1(x,y)\, x$.}
\begin{equation}
   \langle \Delta \lambda_{\max}\rangle = N^{-1/a} f\left(N^{1/a} \Delta \alpha \right)
   \label{eq:scalingLambdaMax}
\end{equation}
The asymptotic limits of the scaling function $f(y)$ are
\begin{equation}
   f(y) \sim 
\begin{cases}
\text{const} & \text{for } y \to -\infty \quad (\text{noise dominated})\\
k y & \text{for } y \simeq 0 \quad (\text{signal onset})
\end{cases}
\end{equation}
The scaling for $y \simeq 0$ is a direct consequence of the linear behaviour in Eq.~\eqref{eq:lambdaoutscaling}, because for $\Delta\alpha \simeq 0$ the maximum eigenvalue is the outlier. Whereas in the $y \to -\infty$ limit, the rank-one perturbation is irrelevant and the maximum eigenvalue scales like $\lambda_1(N)$, see Eq.~\eqref{eq:lambda1scaling}.

To confirm the predicted FSS behavior, we show in Fig.~\ref{fig:scaling_collapse}, for the two ensemble studied, the data for $\langle \lambda_{\max} \rangle$ and the good collapse obtained using Eq.~\eqref{eq:scalingLambdaMax}.

\subsubsection{Statistics of the overlap for the principal eigenvector}

In a discontinuous BBP transition, the overlap $|\bm{v} \cdot \bm{v}_{\max}|$ between the principal eigenvector and the rank-one perturbation has a jump at the critical point $\alpha_\text{BBP}$ in the large $N$ limit. For finite sizes, the mean overlap changes continuously in $\alpha$, and the asymptotic jump becomes a smooth change over the critical region of size $\Delta\alpha \sim N^{-1/a}$.
For these reasons, we assume the following scaling function for the mean overlap
\footnote{A more general FSS assumes for the overlap probability the scaling form $P(|\bm{v}\!\cdot\!\bm{v}_{\max}|;\alpha,N)=H_2\left(|\bm{v}\!\cdot\!\bm{v}_{\max}|,  N^{1/a} \Delta\alpha \right)$ where $H_2(x,y)$ is a scaling function. This ansatz leads to the equation in the main text for $\langle |\bm{v}\!\cdot\!\bm{v}_{\max}| \rangle$ where $g(y)=\int dx\, H_2(x,y)\, x$.}
\begin{equation}
    \langle |\bm{v}\!\cdot\!\bm{v}_{\max}| \rangle = g\left(N^{1/a} \Delta \alpha \right)\,,
    \label{eq:scalingOverlap}
\end{equation}
for $\Delta\alpha \le 0$, where the scaling function $g(y)$ exhibits the following asymptotic behaviors:
\begin{equation*}
    g(y) \sim 
    \begin{cases} 
    |y|^{-a/2} & \text{for } y \to -\infty \quad (\text{noise dominated}) \\
    \text{const} & \text{for } y \to 0 \quad (\text{signal onset})
    \end{cases}
\end{equation*}
For $y \to 0$, the constant value coincides with the overlap value at criticality in the large $N$ limit (which is nonzero in a discontinuous BBP transition).
In the $y \to -\infty$ limit, the rank-one perturbation is irrelevant, and the principal eigenvector is uncorrelated from $\bm{v}$, leading to an overlap of order $N^{-1/2}$, which requires the scaling $g(y) \sim |y|^{-a/2}$.

To verify these predictions, we plot in Fig.~\ref{fig:overlap_collapse}, for the two ensembles studied, the mean overlap as a function of $\Delta \alpha$ (left panels). In the right panels, the data collapse well if scaled according to Eq.~\eqref{eq:scalingOverlap}.

This collapse has a significant physical implication: for any finite system size $N$, the overlap does not vanish abruptly at the thermodynamic transition point $\alpha_{\mathrm{BBP}}$. Instead, there exists a finite region below the theoretical threshold---with a width scaling as $N^{-1/a}$---where the overlap remains finite and distinguishable from the bulk noise floor ($\sim N^{-1/2}$), effectively lowering the detection threshold in finite systems.

\section{Discussion and Conclusions}

In this work, we have shown that the Baik-Ben Arous-P\'ech\'e (BBP) transition---traditionally characterized by the continuous emergence of structure in large noisy matrices---can exhibit a discontinuous behavior, a feature we establish as general across diverse classes of random matrix ensembles. 
We have identified that the main ingredient leading to this new behavior is the decay of the spectral density near the edge. Whenever the eigenvalue density of a random matrix vanishes super-linearly (scaling as $(\lambda_+ - \lambda)^{a-1}$ with $a > 2$), the transition becomes discontinuous. In particular we have illustrated this mechanism for two distinct ensembles: the GOE ensemble with diagonal disorder and the reweighted Wishart ensemble with disordered weights. In both cases, whenever the disordered components induce a faster-than-linear decay in the spectral density, rather than the classical square-root decay, the transition shifts from continuous to discontinuous.

Unlike the standard continuous BBP transition, where signal or structure emerges gradually, the discontinuous BBP transition is characterized by a finite jump at the critical point $\alpha_{\mathrm{BBP}}$ in the overlap $|\bm{v} \cdot \bm{v}_{\mathrm{out}}|$ between the leading eigenvector $\bm{v}_{\mathrm{out}}$ and the hidden structure $\bm{v}$.

Our finite-size scaling analysis shows that for discontinuous transitions, there is a critical region in $\alpha$, \emph{below} the threshold $\alpha_\text{BBP}$, that scales as $N^{-1/a}$ where an outlier is still present and carries a finite amount of the signal. This implies that in finite systems, a detectable signal can emerge from outlier eigenvalues at a signal-to-noise ratio below the large-$N$ prediction.  
Therefore, the asymptotic threshold derived in the thermodynamic limit acts as a conservative upper bound for practical applications, as finite-size fluctuations effectively lower the barrier for signal recovery, allowing structure to emerge in regimes previously considered sub-critical.

Beyond the novel universal random matrix theory framework and the new prominent role of finite-size effects, the relevance of the discontinuous BBP transition is underscored by its potential application across several diverse disciplines. 

We already count two practical implementations of the new mathematical framework of discontinuous BBP transitions in domains as different as phase separation in complex mixtures \cite{thewes2023composition} and inference in over-parametrized neural networks \cite{annesi2025overparametrization}. 

In the first case~\cite{thewes2023composition}, a discontinuous BBP transition has been suggested to emerge in the context of the thermodynamic stability of multi-component biological mixtures, such as the cytoplasm. In these systems, the Hessian of the free energy—which governs phase-ordering spinodal instabilities—can be modeled as a GOE interaction matrix representing the virial coefficients. This matrix is subject to strong diagonal disorder arising from the diverse densities of the available components, alongside a rank-one perturbation linked to the solvent density. The resulting spectral transition defines the relevant spinodal instability and, depending on the distribution of component densities, can manifest as a discontinuous BBP transition.

In the second case~\cite{annesi2025overparametrization}, we turn to a machine learning setting within the framework of two-layer soft-committee machines. Here, one focuses on how over-parameterization, by increasing the number of student hidden units $p$ beyond the task's intrinsic difficulty $p^*$, alters the geometry of the loss landscape. In these settings, it is suggested to analyze the Hessian at random initialization---which has the form of a reweighted Wishart matrix---and use its least eigenvector as a warm start for a subsequent gradient-descent learning algorithm. As the dataset size $\alpha$ increases, the Hessian undergoes a spectral BBP transition. It was observed that increasing over-parameterization ($p>p^*$) not only shifts this transition toward a smaller critical size $\alpha_\text{BBP}$, but fundamentally changes its nature from continuous to discontinuous.
In the large over-parameterization limit, the signal recovery transition always occurs via a discontinuous BBP jump. 

In both cases, the strong finite-size effects we have uncovered must act as a powerful catalyst for the emergence of structure, as there always exists a sub-critical region creating a window where an informative outlier is already present and captures a finite fraction of the signal.
In the biological context, this implies that multi-component mixtures are expected not to need reaching the idealized thermodynamic limit to organize; instead, finite-size fluctuations should allow the system to trigger functional phase-ordering instabilities at concentration imbalances significantly lower than asymptotically predicted.
Analogously, in over-parameterized neural networks, this effect, on the one hand, gives a direct explanation of the large overestimation of expected BBP signal recovery threshold with respect to the numerical results \cite{annesi2025overparametrization}, on the other hand, anticipates the existence of a crucial computational shortcut: by widening the `pre-critical' window, it allows gradient-based optimization to latch onto the hidden signal well before the formal data-requirement threshold is met.

In conclusion, the discovery of the discontinuous BBP transition identifies a new mechanism for the abrupt emergence of signal or structure in large random matrices governed by strong disorder.
In this regime, system finiteness is not a limitation but a fundamental asset. 
By effectively lowering the barrier for signal recovery, finite-size fluctuations transform theoretically ``sub-critical" regimes into fertile ground for the birth of structure. 
This phenomenon suggests that spectral methods in strongly disordered systems may be more powerful---albeit more affected by strong fluctuations---than what the standard continuous theory implies in the large $N$ limit; thus, effectively bridging the gap between abstract random matrix theory and the robust, efficient organization observed in both living matter and artificial intelligence.

The significance of the discontinuous BBP transition extends far beyond the specific examples mentioned here. Much as the classical continuous BBP transition has become a cornerstone for understanding structure detection in high-dimensional data, the discontinuous counterpart can provide a new fundamental framework for systems in areas as diverse as glass physics, socioeconomics, ecology, and inference or data science, particularly when strong disorder is the defining feature—a regime at the forefront of modern statistical physics.
By characterizing the nature of the BBP jump between noise- and signal-dominated regimes in the discontinuous case, including its prominent finite size effects, we provide the new lens through which to analyze the precursors of sharp transitions and informational thresholds in realistic, finite-sized systems, potentially belonging to such a wealth of different areas.

\begin{acknowledgments}
We thank Pierpaolo Vivo and Jean Barbier for useful discussions and insightful comments. We acknowledge financial support from the European Union – NextGenerationEU (PNRR) Fund, Mission 4, Component 2. In particular, we acknowledge support from Investment 1.1 – Project 202234LKBW ``Land(e)scapes: Statistical Physics Theory and Algorithms for Inference and Learning Problems", CUP B53D23003850006 (PRIN 2022). This work was also supported by Investment 1.3 – FAIR Foundation, Extended Partnership “Future Artificial Intelligence Research” (Project Code PE00000013-FAIR), and by Investment 1.4 – National Research Center in “High Performance Computing, Big Data and Quantum Computing” (ICSC, Center Code CN00000013-CN1, CUP B83C22002940006). GB was supported by ANR PRAIRIE-PSAI (France 2023) "ANR-23-IACL-0008". 
\end{acknowledgments}

\bibliography{biblio}

\appendix

\section{Derivation of the Stieltjes Transform for Deformed GOE}
\label{sec:app_deformedGOE_derivation}

In this section, we derive the self-consistent equation for the Stieltjes transform of the deformed GOE ensemble $\bm{M} = \bm{D} + \sigma \bm{W}$ using the framework of free probability \cite{mingo2017free}. In the large $N$ limit, the diagonal matrix $\bm{D}$ and the Wigner matrix $\bm{W}$ become asymptotically free. This allows us to use the property of additivity of the R-transform (the free probability analog of the logarithm of the characteristic function) for the sum of independent non-commutative random variables. The R-transform $R_M(g)$ of the limiting spectral distribution is related to the Stieltjes transform $g(z)$ through the functional equation:
\begin{equation}
    R_M(g) = z_M(g) - \frac{1}{g},
\end{equation}
where $z_M(g)$ is the functional inverse of the Stieltjes transform $g_M(z)$. Using the additivity property $R_{D+\sigma W}(g) = R_D(g) + R_{\sigma W}(g)$, we can express the inverse Stieltjes transform of the sum as:
\begin{equation}
    z(g) - \frac{1}{g} = \left( z_D(g) - \frac{1}{g} \right) + R_{\sigma W}(g),
\end{equation}
where $z_D(g)$ is the inverse Stieltjes transform of the diagonal part $D$. Rearranging the terms, we obtain:
\begin{equation}
    z(g) = z_D(g) + R_{\sigma W}(g).
\end{equation}
For a Wigner matrix $W$ with variance $1/N$, the R-transform is simply $R_W(g) = g$. The scaling factor $\sigma$ transforms the variance to $\sigma^2$, leading to $R_{\sigma W}(g) = \sigma^2 g$. Substituting this into the previous equation yields:
\begin{equation}
    z(g) = z_D(g) + \sigma^2 g.
\end{equation}
This relation implies that $z_D(g) = z - \sigma^2 g$. Since $z_D(g)$ is the inverse function of the Stieltjes transform of $D$, denoted as $g_D(z)$, we can apply $g_D$ to both sides to obtain:
\begin{equation}
    g = g_D(z - \sigma^2 g).
\end{equation}
The Stieltjes transform of the diagonal matrix $D$ is defined by its spectral density $P_D(h)$:
\begin{equation}
    g_D(\omega) = \int \frac{P_D(h)}{\omega - h} dh.
\end{equation}
Substituting $\omega = z - \sigma^2 g$ into this definition recovers the self-consistent equation found in the main text (Eq. 17):
\begin{equation}
    g(z) = \int_{0}^{h_{\max}} \frac{P_D(h)}{z - h - \sigma^2 g(z)} dh.
\end{equation}

\section{Analytical Condition for Discontinuous Transition in Deformed GOE Matrices}
\label{sec:app_deformedGOE_condition}

In this section, we derive the analytical condition under which the BBP transition becomes discontinuous for the deformed GOE ensemble. As established in the general theory, the transition is discontinuous when the spectral density vanishes superlinearly at the edge ($a > 2$) and the inverse Stieltjes transform $z(x)$ reaches the domain boundary without a turning point (i.e., $z'(x) \neq 0$).

For the deformed GOE, the relationship between $z$ and the auxiliary variable $x = g - z/\sigma^2$ is given by:
\begin{equation}
    z(x) = -\sigma^2 \int_{0}^{h_{\max}} dh \, P_D(h) \left[ x + \frac{1}{h + \sigma^2 x} \right].
\end{equation}
The correct branch of solutions is constrained by the condition that the denominator must not vanish, defining the boundary of the domain at $x_c = -h_{\max}/\sigma^2$. A discontinuous transition occurs if $z(x)$ remains monotonic decreasing up to this boundary, which requires $z'(x_c) \le 0$.

Differentiating $z(x)$ with respect to $x$, we obtain:
\begin{equation}
    z'(x) = -\sigma^2 \int_{0}^{h_{\max}} dh \, P_D(h) \left[ 1 - \frac{\sigma^2}{(h + \sigma^2 x)^2} \right].
\end{equation}
Evaluating this at the boundary $x_c = -h_{\max}/\sigma^2$, the term $(h + \sigma^2 x)$ becomes $(h - h_{\max})$. The condition $z'(x_c) \le 0$ leads to:
\begin{equation}
    1 - \sigma^2 \int_{0}^{h_{\max}} \frac{P_D(h)}{(h - h_{\max})^2} \, dh \ge 0.
    \label{eq:app_goe_condition}
\end{equation}
We use the specific distribution for the diagonal elements given in the text:
\begin{equation}
    P_D(h) = \frac{a (h_{\max} - h)^{a-1}}{h_{\max}^a} \quad \text{for } 0 \le h \le h_{\max}.
\end{equation}
Substituting this into the integral term $J$ of Eq.~\eqref{eq:app_goe_condition}:
\begin{align}
    J &= \int_{0}^{h_{\max}} \frac{a (h_{\max} - h)^{a-1}}{h_{\max}^a (h - h_{\max})^2} \, dh\\
    &= \frac{a}{h_{\max}^a} \int_{0}^{h_{\max}} (h_{\max} - h)^{a-3} \, dh.
\end{align}
For $a > 2$, this integral converges. Integrating explicitly:
\begin{align}
    J &= \frac{a}{h_{\max}^a} \left[ \frac{-(h_{\max} - h)^{a-2}}{a-2} \right]_{0}^{h_{\max}}\\
    &= \frac{a}{h_{\max}^a} \left( \frac{h_{\max}^{a-2}}{a-2} \right) = \frac{a}{h_{\max}^2 (a-2)}.
\end{align}
Substituting $J$ back into the condition $1 - \sigma^2 J \ge 0$:
\begin{equation}
    1 - \frac{\sigma^2 a}{h_{\max}^2 (a-2)} \ge 0 \implies \frac{\sigma^2}{h_{\max}^2} \le \frac{a-2}{a}.
\end{equation}
Thus, the analytical condition for the existence of a discontinuous BBP transition in the deformed GOE ensemble is:
\begin{equation}
    \frac{\sigma}{h_{\max}} \le \sqrt{\frac{a-2}{a}}.
\end{equation}

When this condition is satisfied, the upper spectral edge is located at $\lambda_+ = z(x_c)$, where $x_c = -h_{\max}/\sigma^2$. The behavior of the spectral density $\rho(\lambda)$ near this edge can be derived by analyzing the imaginary part of the self-consistent equation for the Stieltjes transform. Recall that the density is given by $\rho(\lambda) = \frac{1}{\pi} \text{Im}[g(\lambda - i\eta)]$ as $\eta \to 0^+$. Looking at the structure of Eq.~\eqref{eq:self_deformedGOE}, the imaginary part arises from the pole in the denominator $z - h - \sigma^2 g(z) = -\sigma^2(x + h/\sigma^2)$. This implies that the spectral density is proportional to the diagonal distribution evaluated at $-\sigma^2 \text{Re}(x)$:
\begin{equation}
    \rho(\lambda) \propto P_D(-\sigma^2 \text{Re}(x)).
\end{equation}
Since the transition is discontinuous, the derivative $z'(x_c)$ is negative and finite. This allows us to expand the inverse Stieltjes transform linearly near the edge. For $z \approx \lambda_+$, we have $z(x) \approx z(x_c) + z'(x_c)(x - x_c)$, which implies:
\begin{equation}
    x \approx x_c + \frac{\lambda - \lambda_+}{z'(x_c)}.
\end{equation}
Substituting this back into the density expression, and noting that $-\sigma^2 x_c = h_{\max}$, we obtain:
\begin{equation}
    \rho(\lambda) \propto P_D\left( h_{\max} - \frac{\sigma^2}{|z'(x_c)|} (\lambda_+ - \lambda) \right).
\end{equation}
Given the power-law tail of the diagonal entries $P_D(h) \sim (h_{\max} - h)^{a-1}$, we conclude that the spectral density of the deformed matrix vanishes superlinearly with the same exponent:
\begin{equation}
    \rho(\lambda) \sim (\lambda_+ - \lambda)^{a-1}.
\end{equation}

\section{Derivation of the Stieltjes Transform for Reweighted Wishart Matrices}
\label{app:app_rewightedWishart_derivation}

In this section, we derive the self-consistent equation for the Stieltjes transform of the reweighted Wishart ensemble defined in Eq. (20). The reweighted Wishart matrix $W$ is defined as the sum of $T$ independent rank-one matrices:
\begin{equation}
    W = \sum_{\mu=1}^{T} X_\mu, \quad \text{where} \quad X_\mu = \frac{f_\mu}{T} \bm{\xi}^\mu (\bm{\xi}^\mu)^\top.
\end{equation}
Here, $\bm{\xi}^\mu$ are random vectors with independent standard Gaussian entries. In the high-dimensional limit where $N, T \to \infty$ with $q = N/T$ fixed, the squared norm of each vector concentrates around its mean: $|\bm{\xi}^\mu|^2 \approx N$. Consequently, each matrix $X_\mu$ is approximately a projector onto a random direction with a single non-zero eigenvalue $\Lambda_\mu$ given by:
\begin{equation}
    \Lambda_\mu \approx \frac{f_\mu}{T} |\bm{\xi}^\mu|^2 \approx \frac{f_\mu N}{T} = q f_\mu.
\end{equation}
In the language of free probability, the matrices $X_\mu$ are asymptotically free independent random variables. The spectral properties of their sum $W$ can be determined again using the additivity of the R-transform:
\begin{equation}
    R_W(g) = \sum_{\mu=1}^{T} R_{X_\mu}(g).
\end{equation}
For a rank-one matrix $X_\mu$ with non-zero eigenvalue $\Lambda_\mu$, the R-transform in the large $N$ limit is given to leading order by:
\begin{equation}
    R_{X_\mu}(g) \approx \frac{1}{N} \frac{\Lambda_\mu}{1 - \Lambda_\mu g}.
\end{equation}
Substituting $\Lambda_\mu = q f_\mu$, we sum over all $T$ components:
\begin{equation}
    R_W(g) = \sum_{\mu=1}^{T} \frac{1}{N} \frac{q f_\mu}{1 - q f_\mu g} = \frac{T}{N} \frac{1}{T} \sum_{\mu=1}^{T} \frac{q f_\mu}{1 - q f_\mu g}.
\end{equation}
Recalling that $T/N = 1/q$, the prefactors cancel ($ \frac{1}{q} \cdot q = 1$), and the sum converges to an expectation over the limiting distribution $P_f(f)$ of the weights:
\begin{equation}
    R_W(g) = \mathbb{E}_f \left[ \frac{f}{1 - q f g} \right].
\end{equation}
Finally, using the functional relation between the R-transform and the inverse Stieltjes transform $z(g) = R_W(g) + 1/g$, we arrive at the result presented in the main text (Eq. 21):
\begin{equation}
    z(g) = \mathbb{E}_f \left[ \frac{f}{1 - q f g} \right] + \frac{1}{g}.
\end{equation}

\section{Analytical Condition for Discontinuous Transition in Reweighted Wishart Matrices}
\label{app:app_rewightedWishart_condition}

In this section, we derive the analytical condition under which the BBP transition becomes discontinuous for the reweighted Wishart ensemble. As discussed in the main text, the transition is discontinuous if the spectral density vanishes superlinearly at the edge ($a>2$) and the following condition on the weights is met:
\begin{equation}
    \mathbb{E}_f \left[ \frac{f^2}{(f_{\max} - f)^2} \right] < q.
    \label{eq:app_condition_start}
\end{equation}
We assume the weight distribution $P_f(f)$ follows a power-law behavior near the edge $f_{\max}$:
\begin{equation}
    P_f(f) = \frac{a (f_{\max} - f)^{a-1}}{f_{\max}^a} \quad \text{for } 0 \le f \le f_{\max}.
\end{equation}
Let $I$ denote the expectation value on the left-hand side of Eq.~\eqref{eq:app_condition_start}. We write the integral explicitly:
\begin{equation}
    I = \int_{0}^{f_{\max}} \frac{f^2}{(f_{\max} - f)^2} \left( \frac{a (f_{\max} - f)^{a-1}}{f_{\max}^a} \right) df.
\end{equation}

Combining the terms with $(f_{\max} - f)$, the integrand simplifies to:
\begin{equation}
    I = \frac{a}{f_{\max}^a} \int_{0}^{f_{\max}} f^2 (f_{\max} - f)^{a-3} \, df.
\end{equation}
We perform the change of variables $u = f/f_{\max}$, and the integral becomes:
\begin{align}
    I &= \frac{a}{f_{\max}^a} \int_{0}^{1} (f_{\max} u)^2 (f_{\max} (1 - u))^{a-3} (f_{\max} du) \nonumber \\
      &= \frac{a}{f_{\max}^a} \cdot f_{\max}^2 \cdot f_{\max}^{a-3} \cdot f_{\max} \int_{0}^{1} u^2 (1 - u)^{a-3} \, du \nonumber \\
      &= a \int_{0}^{1} u^2 (1 - u)^{a-3} \, du.
\end{align}
The powers of $f_{\max}$ cancel out making the condition independent of the cutoff $f_{\max}$. The remaining integral is the definition of the Beta function $\mathcal{B}(x, y) = \int_0^1 u^{x-1}(1-u)^{y-1}du$ with $x=3$ and $y=a-2$:
\begin{equation}
    I = a \, \mathcal{B}(3, a-2).
\end{equation}
For this integral to converge, the exponent of the singularity must satisfy $a-3 > -1$, which recovers the necessary condition $a > 2$ for the existence of a superlinear edge. 

Using the relationship between the Beta and Gamma functions, $\mathcal{B}(x,y) = \frac{\Gamma(x)\Gamma(y)}{\Gamma(x+y)}$, we obtain:
\begin{equation}
    I = a \frac{\Gamma(3)\Gamma(a-2)}{\Gamma(3 + a - 2)} = a \frac{2! \, \Gamma(a-2)}{\Gamma(a+1)}.
\end{equation}

Using the property of the Gamma function, we expand the denominator as $\Gamma(a+1) = a(a-1)(a-2)\Gamma(a-2)$. This leads to the final compact expression:
\begin{equation}
    I = \frac{2 a \Gamma(a-2)}{a(a-1)(a-2)\Gamma(a-2)} = \frac{2}{(a-1)(a-2)}.
\end{equation}
Substituting this back into Eq.~\eqref{eq:app_condition_start}, the condition for a discontinuous transition (assuming $a>2$) becomes:
\begin{equation}
    q > \frac{2}{(a-1)(a-2)}.
\end{equation}
This inequality determines the phase boundary in the $(a, q)$ plane where the transition switches from continuous to discontinuous, independently of the parameter $f_{\max}$. For $a=3$, this requires $q > 1$.

When this condition is satisfied, the upper spectral edge is located at $\lambda_+ = z(g_{\max})$, where $g_{\max} = (q f_{\max})^{-1}$. The behavior of the spectral density $\rho(\lambda)$ near this edge can be derived by analyzing the imaginary part of the self-consistent equation for the Stieltjes transform $g(z)$. For $z = \lambda + i \eta$ with $\eta \to 0^+$, the imaginary part of the relation $z(g) = z$ yields:
\begin{equation}
    \frac{\text{Im}(g)}{|g|^2} \approx \pi \mathbb{E}_{f} [\delta(1 - q f \text{Re}(g))] \implies \rho(\lambda) \propto P_f \left( \frac{1}{q \text{Re}(g)} \right).
\end{equation}
Since the transition is discontinuous, the derivative $z'(g_{\max})$ is negative and finite. This allows us to expand the Stieltjes transform linearly near the edge \begin{equation}
    g(\lambda) \approx g_{\max} + \frac{\lambda - \lambda_+}{z'(g_{\max})}.
\end{equation}
Substituting this into the density expression leads to:
\begin{equation}
    \rho(\lambda) \propto P_f \left( f_{\max} - \frac{f_{\max}}{g_{\max} |z'(g_{\max})|} (\lambda_+ - \lambda) \right).
\end{equation}
Given the power-law tail of the weights $P_f(f) \sim (f_{\max} - f)^{a-1}$, we conclude that the spectral density vanishes superlinearly with the same exponent:
\begin{equation}
    \rho(\lambda) \sim (\lambda_+ - \lambda)^{a-1}.
\end{equation}

\section{Finite-$N$ continuation of the outlier eigenvalue in the deformed GOE ensemble}
\label{sec:app_finiteN_GOE_outlier}

In this section, we explain more precisely why, in the deformed GOE ensemble, the outlier eigenvalue can be continued in $\alpha$ below the thermodynamic transition point when $N$ is large but finite, and why in that regime it still satisfies the usual overlap formula and the linearization used in the main text.

We consider the spiked deformed GOE matrix
\begin{equation}
    \bm{M}_\alpha = \bm{A} + \alpha \bm{v}\bm{v}^\top,
\end{equation}
where
\begin{equation}
    \bm{A} = \bm{D} + \sigma \bm{W},
\end{equation}
and $\bm{v}$ is independent of $\bm{A}$ and has unit norm. Let $\{\lambda_a^0,\bm{u}_a^0\}$ be the eigenvalues and eigenvectors of $\bm{A}$, ordered so that
\begin{equation}
\lambda_1^0\ge \lambda_2^0\ge \cdots \ge \lambda_N^0,
\end{equation}
and let $\{\lambda_a(\alpha),\bm{u}_a(\alpha)\}$ be the eigenvalues and eigenvectors of $\bm{M}_\alpha$, ordered so that
\begin{equation}
\lambda_1(\alpha)\ge \lambda_2(\alpha)\ge \cdots \ge \lambda_N(\alpha).
\end{equation}

The Sherman--Morrison formula can be used to extract the spectrum of any matrix $\bm{A}-\bm{B}$ from the knowledge of the spectrum of $\bm{A}$, provided that $\bm{B}$ is a rank-one correction, $\bm{B}=-\alpha \bm{v}\bm{v}^T$. The Sherman--Morrison formula for the inverse of the sum of a matrix $\bm{A}$ and a rank-one matrix $\bm{B}=-\alpha \bm{v}\bm{v}^T$ is
\begin{equation}
    (\bm{A}+\bm{B})^{-1}
    =
    \bm{A}^{-1}
    -
    \frac{\bm{A}^{-1}\bm{B}\bm{A}^{-1}}{1+\mathrm{Tr}(\bm{B}\bm{A}^{-1})}.
\end{equation}

To apply it to the eigenvalue problem, $\bm{A}^{-1}$ is identified with the resolvent $\bm{G}_0(z)=(z\bm{I}-\bm{A})^{-1}$ of the matrix $\bm{A}$, such that
\begin{equation}
\mathrm{Tr}\,\bm{G}_0(z)=\sum_a\frac{1}{z-\lambda_a^0},
\end{equation}
and $(\bm{A}+\bm{B})^{-1}$ with the resolvent $\bm{G}_{\alpha}(z)=(z\bm{I}-\bm{A}-\bm{B})^{-1}$ of $\bm{M}_\alpha$ so that
\begin{equation}
\mathrm{Tr}\,\bm{G}_{\alpha}(z)=\sum_a\frac{1}{z-\lambda_a(\alpha)}.
\end{equation}

The relation of the two resolvents is then
\begin{equation}
    \bm{G}_{\alpha}(z)
    =
    \bm{G}_0(z)
    +
    \alpha\frac{\bm{G}_0(z)\bm{v}\bm{v}^T\bm{G}_0(z)}{1-\alpha\bm{v}^T\bm{G}_0(z)\bm{v}}.
\end{equation}

In the eigenbasis $\{\bm{u}_a^0\}$ where $\bm{A}$ is diagonal with elements $\lambda_a^0$, we have
\begin{equation}
\mathrm{Tr}(\bm{G}_0(z)\bm{v}\bm{v}^T\bm{G}_0(z))
=
\sum_a\frac{(\bm{u}_a^0\!\cdot\!\bm{v})^2}{(z-\lambda_a^0)^2},
\end{equation}
and
\begin{equation}
\bm{v}^T\bm{G}_0(z)\bm{v}
=
\sum_a\frac{(\bm{u}_a^0\!\cdot\!\bm{v})^2}{z-\lambda_a^0}.
\end{equation}

Therefore
\begin{equation}
    \mathrm{Tr}(\bm{G}_{\alpha}(z))
    =
    \sum_a\frac{1}{z-\lambda_a^0}
    +
    \alpha
    \frac{\sum_a\frac{(\bm{u}_a^0\cdot\bm{v})^2}{(z-\lambda_a^0)^2}}
    {1-\alpha\sum_a\frac{(\bm{u}_a^0\cdot\bm{v})^2}{z-\lambda_a^0}}.
\end{equation}

Note that, when $z\rightarrow \lambda_b^0$, only one term dominates the sums and
\begin{equation}
\mathrm{Tr}(\bm{G}_{\alpha}(z\rightarrow \lambda_b^0))
\sim
\frac{1}{z-\lambda_b^0}
+
\alpha
\frac{\frac{(\bm{u}_b^0\cdot\bm{v})^2}{(z-\lambda_b^0)^2}}
{1-\alpha\frac{(\bm{u}_b^0\cdot\bm{v})^2}{z-\lambda_b^0}}
\sim 0.
\end{equation}
Due to this cancellation, the poles of $\mathrm{Tr}(\bm{G}_0(z))$ do not correspond exactly to the poles of $\mathrm{Tr}(\bm{G}_{\alpha}(z))$. The latter are instead defined by the points where the second term diverges, or equivalently by the solutions of the equation
\begin{equation}
   \alpha^{-1}= \bm{v}^T\bm{G}_0(z)\bm{v}.
   \label{eq:neweigenGOE}
\end{equation}

Starting from Eq.~\eqref{eq:neweigenGOE}, one can also derive the overlap between the eigenvector associated with a solution $\lambda(\alpha)$ and the spike direction $\bm{v}$. Let $\bm{u}(\alpha)$ be a normalized eigenvector of $\bm{M}_\alpha$ associated with $\lambda(\alpha)$. From
\begin{equation}
(\bm{A}+\alpha \bm{v}\bm{v}^T)\bm{u}(\alpha)=\lambda(\alpha)\bm{u}(\alpha),
\end{equation}
we obtain
\begin{equation}
(\lambda(\alpha)\bm{I}-\bm{A})\bm{u}(\alpha)
=
\alpha\,\bm{v}\,(\bm{v}^T\bm{u}(\alpha)),
\end{equation}
hence
\begin{equation}
\bm{u}(\alpha)
=
\alpha\,\bm{G}_0(\lambda(\alpha))\,\bm{v}\,(\bm{v}^T\bm{u}(\alpha)).
\end{equation}

Imposing normalization, one finds
\begin{equation}
1
=
\alpha^2(\bm{v}^T\bm{u}(\alpha))^2
\,\bm{v}^T\bm{G}_0(\lambda(\alpha))^2\bm{v},
\end{equation}
which yields
\begin{equation}
|\bm{v}\!\cdot\!\bm{u}(\alpha)|^2
=
\frac{1}{\alpha^2\,\bm{v}^T\bm{G}_0(\lambda(\alpha))^2\bm{v}}.
\end{equation}

Using
\begin{equation}
\frac{d}{dz}\big[\bm{v}^T\bm{G}_0(z)\bm{v}\big]
=
-\,\bm{v}^T\bm{G}_0(z)^2\bm{v},
\end{equation}
we obtain the expression
\begin{equation}
|\bm{v}\!\cdot\!\bm{u}(\alpha)|^2
=
-\frac{1}{\alpha^2\,\partial_z\!\left[\bm{v}^T\bm{G}_0(z)\bm{v}\right]_{z=\lambda(\alpha)}}.
\label{eq:finiteN_overlap_exact}
\end{equation}

We now use the exact finite-$N$ results on the upper spectral edge of the deformed GOE. As proved in \cite{lee2016extremal}, when the density of states vanishes as
\begin{equation}
\rho_0(\lambda)\sim (\lambda_+-\lambda)^{a-1},
\qquad a>2,
\end{equation}
the largest eigenvalues of $\bm{A}$ have Weibull statistics. In particular, the largest eigenvalue $\lambda_1^0$ lies at distance of order $N^{-1/a}$ from the asymptotic edge $\lambda_+$, and the spacing between the first $K\ll N$ eigenvalues is also of order $N^{-1/a}$.

When $z$ is far from the set of $\lambda_a^0$ and in particular for $z>\lambda_1^0$, we can approximate the right-hand side of Eq.~\eqref{eq:neweigenGOE} by its large-$N$ expression
\begin{equation}
   \bm{v}^T\bm{G}_0(z)\bm{v}
   \simeq
   \sum_a\frac{1}{N(z-\lambda_a^0)}
   \simeq
   \int d\lambda \frac{\rho_0(\lambda)}{z-\lambda}
   \equiv g(z),
   \label{eq:largeNGOE}
\end{equation}
where we used $(\bm{u}_a^0\!\cdot\!\bm{v})^2\simeq 1/N$ and $\rho_0(\lambda)\simeq\frac{1}{N}\sum_a\delta(\lambda-\lambda_a^0)$ is the density of eigenvalues of $\bm{A}$.

The solution of Eq.~\eqref{eq:neweigenGOE}, $\lambda^*(\alpha)>\lambda_+$, represents the usual isolated BBP eigenvalue for $\alpha>\alpha_{\mathrm{BBP}}$. In the infinite-$N$ limit, no real solution can be continued below $\alpha_{\mathrm{BBP}}$, because once $z$ enters the support of the spectrum the thermodynamic Stieltjes transform is no longer real. At finite $N$, however, the spectrum is discrete, and between two consecutive eigenvalues there are real intervals where $\bm{v}^T\bm{G}_0(z)\bm{v}$ is perfectly well defined.

This is the crucial point in the discontinuous case. Since the poles in the discrete sum dominate only on a scale $N^{-1}$, while the distance of $\lambda_1^0$ from $\lambda_+$ and the spacing of the top eigenvalues are of order $N^{-1/a}$, and since for $a>2$
\begin{equation}
N^{-1/a}\gg N^{-1},
\end{equation}
there exists, for large but finite $N$, an extended window below the asymptotic edge where one can move away from the individual poles while still staying above the largest eigenvalue of the unperturbed matrix:
\begin{equation}
\lambda_1^0 < z < \lambda_+.
\end{equation}

In that region, one can approximate $\bm{v}^T\bm{G}_0(z)\bm{v}$ by the sum of a singular contribution from the nearest pole and a smooth principal-value part:
\begin{equation}
   \bm{v}^T\bm{G}_0(z)\bm{v}
   \simeq
   \frac{1}{N(z-\lambda_a^0)}
   +
   \mathrm{P.V.} \int d\lambda \frac{\rho_0(\lambda)}{z-\lambda}.
   \label{eq:finiteNGOE}
\end{equation}
The first term matters only within a distance $N^{-1}$ from the pole $\lambda_a^0$, whereas throughout the whole window $\lambda_1^0<z<\lambda_+$ the second term varies on the much larger scale $N^{-1/a}$. Therefore, in the discontinuous case, there is a whole pre-critical region where the pole term is subleading to leading order and one can still use the same equations as in the thermodynamic limit:
\begin{equation}
\bm{v}^T\bm{G}_0(z)\bm{v}\simeq g(z).
\end{equation}

As a consequence, Eq.~\eqref{eq:neweigenGOE} admits in this region a distinguished solution $\lambda^*(\alpha)$ such that
\begin{equation}
\lambda_1^0<\lambda^*(\alpha)<\lambda_+.
\end{equation}
This solution is the finite-$N$ continuation of the outlier eigenvalue below the asymptotic threshold. It is precisely in this interval that the outlier exists even for $\alpha<\alpha_{\mathrm{BBP}}$.

Most of the other solutions of Eq.~\eqref{eq:neweigenGOE} remain instead very close to the original eigenvalues $\lambda_a^0$, and therefore inherit their large sample-to-sample fluctuations of order $N^{-1/a}$. The distinguished solution $\lambda^*(\alpha)$ is special because it is controlled by the smooth principal-value contribution rather than by a nearby pole. As a result, it can be identified by its much weaker fluctuations and by the fact that its eigenvector remains correlated with the signal.

Since in the controlled region $\lambda_1^0<\lambda^*(\alpha)<\lambda_+$ Eq.~\eqref{eq:finiteN_overlap_exact} reduces to the thermodynamic expression, the overlap is still given by
\begin{equation}
|\bm{v}\!\cdot\!\bm{u}^*(\alpha)|^2
\simeq
-\frac{1}{\alpha^2 g'(\lambda^*(\alpha))},
\label{eq:finiteN_overlap_thermo}
\end{equation}
which coincides with Eq.~\eqref{eq:overlap} in the main text. In particular, because $g'(\lambda_+)$ is finite in the discontinuous case, the overlap remains finite as the outlier eigenvalue crosses below $\lambda_+$ at finite $N$.\\

\begin{figure}[h!]
    \centering
    \vspace{0.0cm}
    \includegraphics[width=1.0\linewidth]{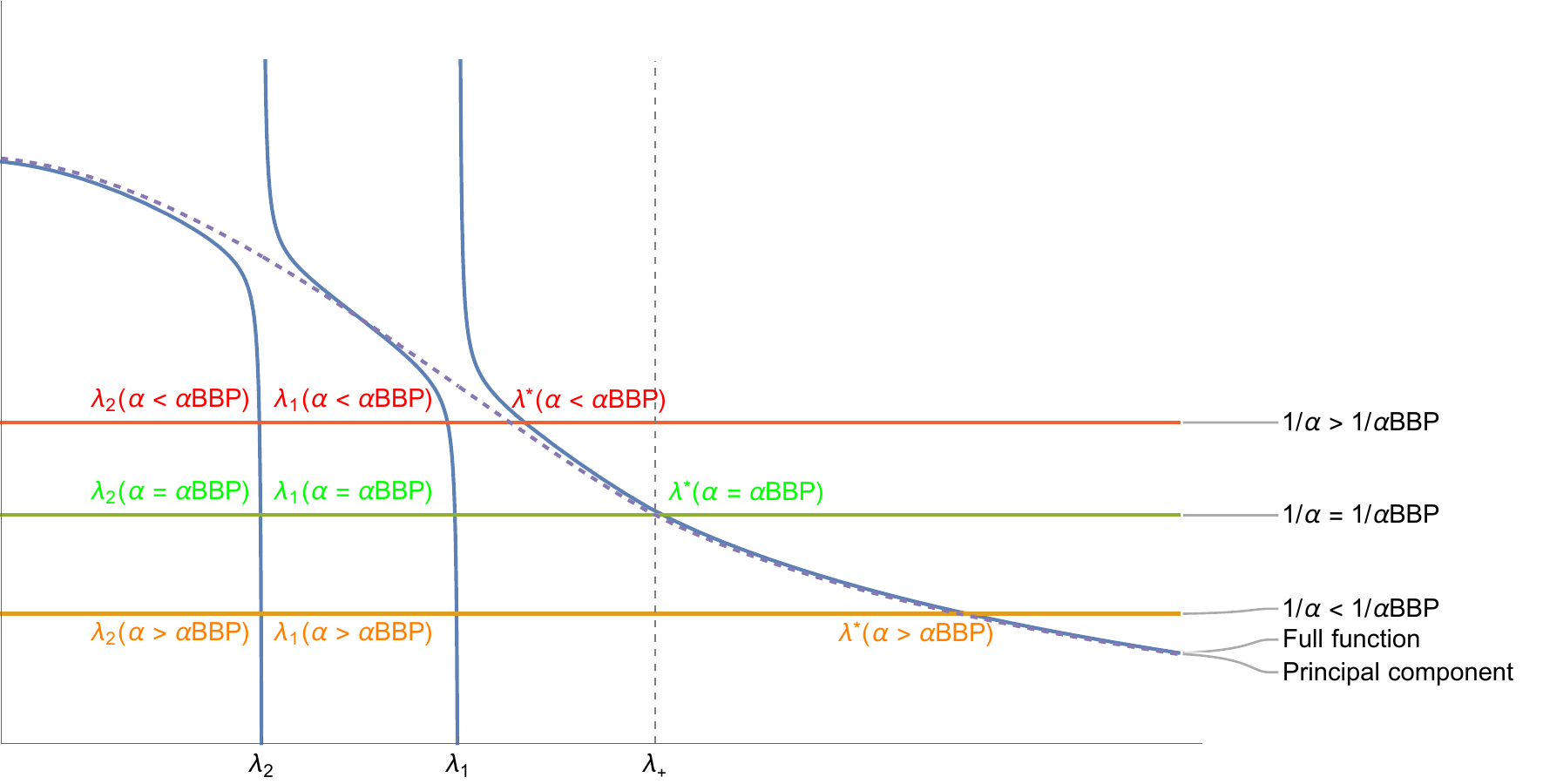}
    \caption{
    Qualitative behavior of ${\bf v}^{T}G_0(z){\bf v}$ near the upper edge. 
    The horizontal lines show the equation ${\bf v}^{T}G_0(z){\bf v}=1/\alpha$ for three regimes: 
    $\alpha>\alpha_{\rm BBP}$, where the outlier lies above the asymptotic edge $\lambda_+$; 
    $\alpha=\alpha_{\rm BBP}$, where it coincides with $\lambda_+$; 
    and $\alpha<\alpha_{\rm BBP}$, where at finite $N$ it can lie below $\lambda_+$ but still above the largest eigenvalue $\lambda_1$.
    }
    \label{fig:chiarafigure}
\end{figure}

Figure \ref{fig:chiarafigure} shows that, depending on $\alpha$, the distinguished solution obtained from the smooth principal-value part may in fact continue also below $\lambda_1^0$, in which case it is no longer the largest eigenvalue of the sample and becomes embedded among the top bulk eigenvalues of the spiked matrix. A full characterization of this regime, and a precise criterion to identify it at finite $N$, is left for future work. Here we restrict ourselves to the controlled region
\begin{equation}
\lambda_1^0<\lambda^*(\alpha)<\lambda_+,
\end{equation}
for which the above argument is explicit.

Finally, near the BBP point one can linearize the equation
\begin{equation}
g(\lambda^*(\alpha))=\alpha^{-1}
\end{equation}
around $(\alpha_{\mathrm{BBP}},\lambda_+)$. Expanding to first order gives
\begin{equation}
\alpha^{-1}-\alpha_{\mathrm{BBP}}^{-1}
\simeq
g'(\lambda_+)\,(\lambda^*(\alpha)-\lambda_+),
\end{equation}
hence
\begin{align}
\lambda^*(\alpha)-\lambda_+
&\simeq
-\frac{1}{\alpha_{\mathrm{BBP}}^2 g'(\lambda_+)}\,(\alpha-\alpha_{\mathrm{BBP}})
\\
&=
m^2(\alpha_{\mathrm{BBP}})\,(\alpha-\alpha_{\mathrm{BBP}}).
\label{eq:finiteN_linearized_outlier}
\end{align}

This shows that, for large but finite $N$, there exists a pre-critical region in which the outlier eigenvalue survives below $\alpha_{\mathrm{BBP}}$, with
\begin{equation}
\lambda_1^0<\lambda^*(\alpha)<\lambda_+,
\end{equation}
and in that region the outlier remains informative and obeys the same overlap formula and linearization as in the thermodynamic theory.
\end{document}